\documentclass[journal]{IEEEtran}

\usepackage[pdftex]{graphicx}
\graphicspath{{../pdf/}{../jpeg/}}
\DeclareGraphicsExtensions{.pdf,.jpeg,.png}
\usepackage[table,xcdraw]{xcolor} 

\usepackage{amsmath,amsfonts} 
\usepackage{array}
\usepackage{eqparbox}

\usepackage{booktabs}  
\usepackage{makecell}  
\usepackage{multirow}  
\usepackage{tabularx}  

\usepackage{algorithm}
\usepackage{algorithmic}

\usepackage[caption=false,font=normalsize,labelfont=sf,textfont=sf]{subfig}

\usepackage{textcomp}
\usepackage{stfloats}
\usepackage{url}
\usepackage{verbatim}
\usepackage{cite}     
\usepackage{comment}
\usepackage{hyperref} 

\usepackage{soul,framed}
\colorlet{shadecolor}{yellow}

\begin{document}


\title{MTSIC: Multi-stage Transformer-based GAN for Spectral Infrared Image Colorization}

\author{
Tingting Liu, Yuan Liu, Jinhui Tang, Liyin Yuan,\\ 
Chengyu Liu, Chunlai Li, Xiubao Sui, and Qian Chen%
\thanks{Tingting Liu is with the School of Electronic Engineering and Optoelectronic Technology, Nanjing University of Science and Technology, Nanjing 210094, China (e-mail: tingtingliu@njust.edu.cn).}%
\thanks{Corresponding authors: Yuan Liu and Xiubao Sui.}
}

\maketitle

\begin{abstract}
Thermal infrared (TIR) images, acquired through thermal radiation imaging, are unaffected by variations in lighting conditions and atmospheric haze. However, TIR images inherently lack color and texture information, limiting downstream tasks and potentially causing visual fatigue. Existing colorization methods primarily rely on single-band images with limited spectral information and insufficient feature extraction capabilities, which often result in image distortion and semantic ambiguity. In contrast, multiband infrared imagery provides richer spectral data, facilitating the preservation of finer details and enhancing semantic accuracy. In this paper, we propose a generative adversarial network (GAN)-based framework designed to integrate spectral information to enhance the colorization of infrared images. The framework employs a multi-stage spectral self-attention Transformer network (MTSIC) as the generator. Each spectral feature is treated as a token for self-attention computation, and a multi-head self-attention mechanism forms a spatial-spectral attention residual block (SARB), achieving multi-band feature mapping and reducing semantic confusion. Multiple SARB units are integrated into a Transformer-based single-stage network (STformer), which uses a U-shaped architecture to extract contextual information, combined with multi-scale wavelet blocks (MSWB) to align semantic information in the spatial-frequency dual domain. Multiple STformer modules are cascaded to form MTSIC, progressively optimizing the reconstruction quality. Experimental results demonstrate that the proposed method significantly outperforms traditional techniques and effectively enhances the visual quality of infrared images. 
\end{abstract}


%
\IEEEpeerreviewmaketitle

\section{Introduction}

\IEEEPARstart{T}{hermal} infrared (TIR) imaging captures objects' thermal radiation, enabling all-weather applications in areas such as security surveillance, automotive navigation, and nighttime traffic monitoring \cite{[2]}. Unlike visible-light images, TIR images are typically grayscale, lacking both color and fine texture details \cite{[33]}. The human visual system can discern thousands of hues and intensities, but only around two dozen shades of gray \cite{coclor_gray}. Prolonged viewing of grayscale images can also lead to visual fatigue, further highlighting the necessity of colorization. Colorization enhances visual expressiveness, improves information conveyance, and facilitates human-computer interaction. Much like the historical shift from black-and-white to color film, colorized images are better suited for object recognition, interpretation, and analysis. Colorization involves assigning colors to grayscale images, yet a single gray value can correspond to a wide range of plausible colors depending on the real-world context. For instance, a flower might appear red, yellow, or white, while a car could have numerous possible color schemes. As a result, the colorization task is inherently nonlinear and ill-posed \cite{MUGAN}. An effective colorization does not require exact replication of the original colors but should instead appear natural and perceptually convincing \cite{BiSTNet}. The goal of natural colorization is to produce realistic and credible color distributions rather than a single "ground truth." Despite the subjective nature of this process, it must adhere to semantic constraints—for example, skies should not appear purple, nor should deserts be colored blue.

TIR image colorization is an image-to-image translation task that aims to learn an optimal mapping from the TIR domain (source) to the color visible domain (target) \cite{[16]}. This task is particularly challenging due to fundamental differences in imaging mechanisms between IR and visible-light modalities. While TIR images encode thermal radiation as grayscale intensities, they lack chromatic information and often exhibit reduced structural detail compared to visible-light images \cite{[18]}. Existing colorization methods designed for visible grayscale images are often inadequate for TIR data. For instance, in nighttime TIR images, occupants inside vehicles may blend into the background due to uniform thermal signatures, whereas visible grayscale images—even under low illumination—retain some structural cues. Moreover, even with pixel-aligned visible-light reference images, reconstructing missing chromatic and textural details during TIR colorization remains difficult \cite{[18]}. Thus, TIR colorization fundamentally requires disentangling content from appearance, requiring both the enhancement of texture and semantic consistency.

An ideal TIR image colorization should produce natural colors while maintaining semantic consistency and sharp structural edges, which is crucial for downstream vision tasks. The colorization process typically consists of three stages—encoding, mapping, and rendering—each presenting distinct challenges \cite{[31]}. In the encoding stage, due to the absence of color and textural cues in IR images, the network may suffer from neighborhood entanglement, where semantically distinct regions (e.g., vehicles and road surfaces) are incorrectly grouped due to similar thermal signatures. During mapping stage, constrained by semantic supervision, the model may oversimplify textures to fit broad categorical priors, leading to semantic distortion. This issue is particularly pronounced in small-sample categories, where limited training data exacerbates feature misalignment. In the rendering stage, small objects are prone to appearance distortion, as their limited pixel area and complex features make colorization unstable. This results in reduced naturalness and visual coherence.

Traditional colorization methods, such as manual coloring, lookup tables, histogram matching, and reference image fusion, rely on prior knowledge and struggle to adaptively align infrared images with semantics and colors \cite{Traditional_colorization}. Recent advances in deep learning have introduced new approaches, primarily CNN- and GAN-based methods \cite{[23]}. CNNs achieve colorization through cross-domain mapping and pixel-level losses but often suffer from blurring, color shifts, and detail loss. GANs employ adversarial loss to constrain appearance and enhance detail expression with pixel loss. Although these methods continuously optimize architectures and strategies, performing well in simple scenes, they still face challenges in complex environments, such as semantic inconsistency, difficulty in small-object recognition, and feature confusion. The main reasons include: (1) limited model expressiveness, and (2) reliance on single-band input, leading to an information bottleneck that hinders compensation for source data deficiencies. In contrast, infrared spectral imaging captures radiation characteristics across multiple narrow bands, enhancing spectral dimensionality and information richness from the source. It provides stronger discriminative ability for materials, boundaries, and environments, facilitating realistic color restoration, semantic structure preservation, and improved interpretability \cite{MBNet}. Therefore, enhancing colorization performance depends not only on network design but also on improving input image quality. Spectral images, richer in details and semantics than single-band images, are widely used in road segmentation, environmental monitoring, and other tasks \cite{[21],[22],[36],[37]}.

However, due to the high dimensionality and strong inter-band correlation of spectral images, existing methods designed for single-band images fail to effectively model spectral dependencies, limiting their ability to fully extract informative features. High-dimensional feature extraction requires deeper networks, yet mainstream GAN generators often adopt UNet \cite{[29],[16]}, ResNet \cite{[62],[31]}, or their variants \cite{[18],[23],MUGAN}, which have limited representational capacity and tend to produce blurred or distorted textures. Moreover, with insufficient training samples, spectral images are prone to the “Hughes phenomenon,” leading to degraded model performance \cite{[22],MBNet,BGSR}. Thus, effectively modeling spatial-spectral information and optimizing networks to improve colorization accuracy remain key challenges.

\begin{figure}[htbp]
    \centering
    \includegraphics[scale=0.40, trim=0 0 0 0, clip]{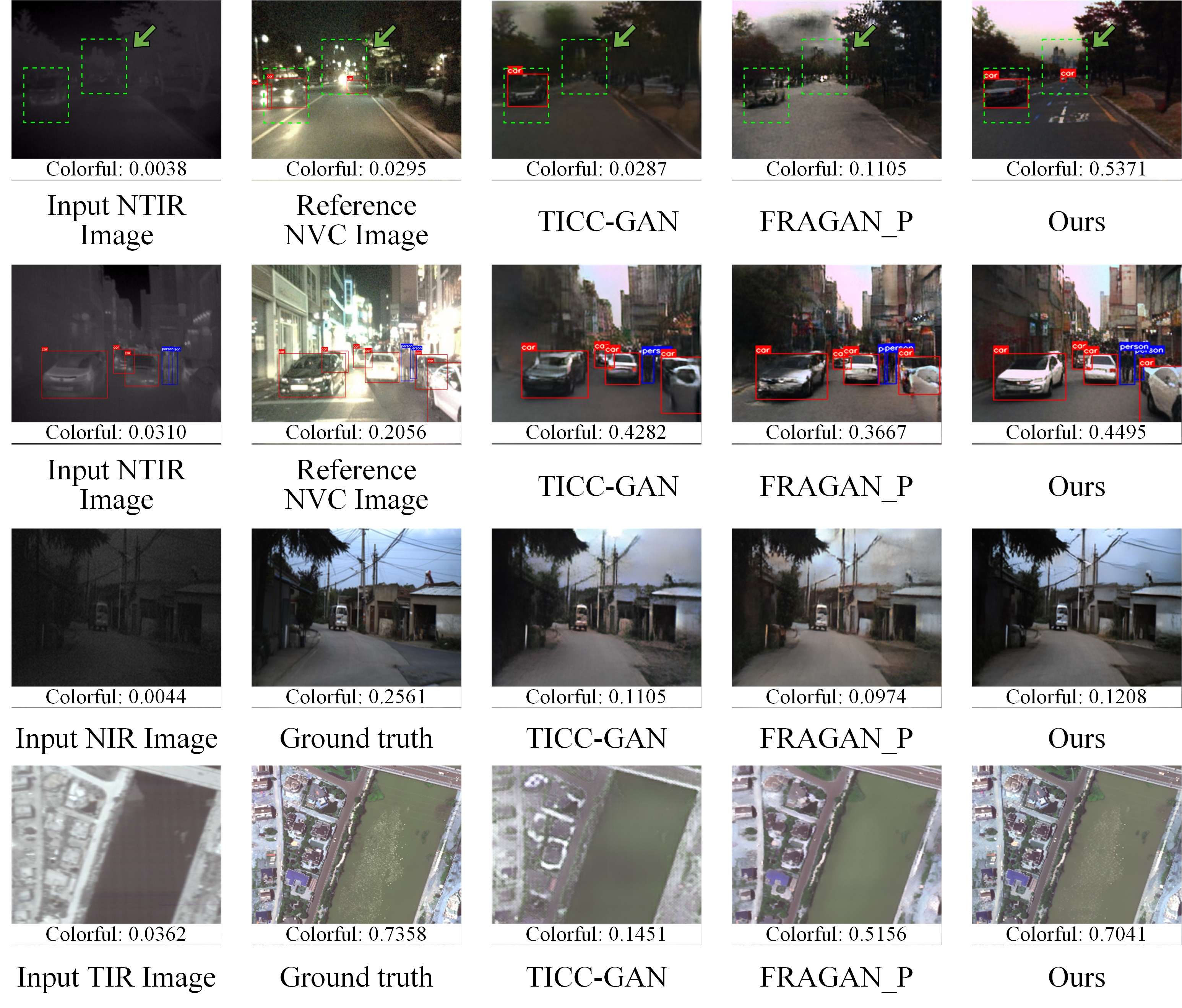}
    \caption{A visual comparison of infrared image colorization results is presented. NTIR denotes nighttime thermal infrared multispectral images, and NIR refers to near-infrared hyperspectral images. NVC represents the reference night-time visible color image. The first two rows show object detection results using YOLOv7 \cite{[19]} trained on the MS COCO dataset \cite{[20]}.}
    \label{fig:Frist_comparison}
\end{figure}

To address these issues, this paper proposes a conditional GAN framework named MTSIC from two perspectives: spectral information enhancement and structural optimization. On one hand, multi-band infrared spectral images are utilized as input, expanding the input from a single spectral dimension to a higher-dimensional representation, which significantly enriches semantic perception and detail recovery. On the other hand, the network structure is optimized to improve feature extraction and mapping capabilities. Specifically, considering the spatial sparsity and high inter-band correlation of spectral images, a multi-head self-attention mechanism along the spectral dimension is introduced in the generator to enable collaborative modeling of multi-band features via query tokens. A single-stage spectral Transformer (STformer) is constructed using the SARB module as the basic unit. The STformer adopts a U-shaped architecture to extract contextual information, effectively alleviating detail loss and semantic confusion in complex backgrounds. To enhance colorization of small target regions, a multi-scale wavelet transform is integrated into the bottleneck layer, achieving cross-scale alignment in both frequency and spatial domains. Multiple STformers are cascaded to form the MTSIC network for progressive reconstruction refinement. In addition, frequency, edge, and spectral angle losses are incorporated into the conventional GAN loss to improve reconstruction quality in terms of texture, structure, and color. 

As shown in \autoref{fig:Frist_comparison}, the proposed method outperforms two classical single-band infrared colorization approaches in visual quality and semantic preservation. The first two subfigures show YOLOv7 detection results on the multispectral KAIST dataset. In the first, the thermal image lacks texture detail, making targets hard to detect, while the NVC image introduces false detections due to complex lighting (e.g., unclear vehicles, misclassified roads). By enhancing color and structure, the proposed method significantly improves detection. The second subfigure shows better preservation of building geometry, effectively reducing structural distortion.  

In summary, the main contributions of this work are outlined as follows:

(1) A new GAN framework incorporating spectral information is proposed, marking the first deep learning-based approach for natural colorization of infrared hyperspectral images. 

(2) A novel multi-stage Transformer-based generator, MTSIC, is proposed. At each stage, a spectral multi-head self-attention mechanism captures global spectral correlations, integrating spatial-spectral features to reduce semantic confusion and improve colorization performance.  

(3) A multiscale wavelet transform is incorporated into the single-stage Transformer for frequency-domain feature mapping, enabling semantic alignment between frequency and spatial-domain features. This integration allows the model to better focus on regions with significant edge and texture variations. 

(4) The infrared hyperspectral remote sensing dataset used for colorization was self-collected. In addition, multiple evaluation metrics were introduced to comprehensively validate the effectiveness of the proposed method.

\section{Related Work}

\subsection{Existing Methods for Infrared Image Colorization}

Infrared images often suffer from detail loss and low quality, making conventional grayscale colorization methods ineffective. Specialized models are needed to improve feature extraction and reconstruction. Current approaches fall into two categories: CNN-based and GAN-based methods \cite{[23]}. CNNs, such as lightweight UNet variants like TIR \cite{[27]}, are efficient for simple scenes but lack detail recovery. Limmer et al. \cite{[28]} introduced high-frequency feature transfer to preserve some details, but performance remains limited. GAN-based methods, leveraging strong generative capabilities, are widely adopted. Kuang et al. \cite{[16]} enhanced Pix2Pix \cite{[29]} with improved generators and loss functions to boost realism, but small target details were lost. Chen et al. \cite{[18]} and Sigillo et al. \cite{[30]} refined TICC-GAN \cite{[16]} to enhance edges, yet unrealistic colors and texture loss persist. Lou et al.'s MornGAN \cite{[31]} uses a ResNet generator and memory guidance to reduce feature aliasing. However, its segmentation loss relies heavily on the performance of the segmentation network, and the generated colors are unrealistic. He et al. \cite{[13]} combined UNet with ViT for better semantic decoding, though small target confusion remained. Chen et al. \cite{[23]} improved generator (UNet++) in FRAGAN for better detail and semantics, but small target colorization and structure reconstruction are still limited. Liao et al.’s MUGAN \cite{MUGAN}, merging UNet++ and UNet3+ variants as generator, also struggles with small target loss and geometric distortion in complex scenes.

Overall, despite progress in deep learning for infrared image colorization, two main challenges remain: (1) semantic confusion and unrealistic colors in cross-domain colorization, which hinder accurate small target recognition and content consistency; (2) reliance on single-band images and focus on vehicular scenes, limiting effectiveness in complex scenarios like remote sensing. As infrared images are widely used in remote sensing tasks such as object extraction, change detection, and environmental monitoring \cite{[33]}, integrating multi-band information is essential to improve colorization quality and address current limitations.

\subsection{Infrared spectral image colorization}

Multi-band infrared spectral images offer richer radiative information, higher material identification accuracy, and better adaptability than single-band images \cite{[21]}. By capturing responses across multiple narrow bands, they enable precise identification, especially for distinguishing similar hetero-isomers, and are widely used in applications like driver assistance and road segmentation \cite{[35]}.

Colorizing multi-band infrared images enhances visualization and detail perception. TeX Vision, introduced in HADAR \cite{[37]}, generates pseudo-color images in HSV space using a thermal signal equation based on temperature (T), emissivity (e), and texture (X). However, the multi-to-one mapping from thermal radiation (S) to T/e/X makes the problem ill-posed \cite{[38]}, reducing recognition accuracy in complex scenes and for small targets. TeX Vision also depends on pre-built material libraries, which may not generalize to new materials, and suffers from high computational complexity and sensitivity to environmental factors like temperature and humidity.

To address these issues, this paper proposes a deep learning-based, data-driven method that learns semantic and color mappings between multi-band infrared and visible images to generate natural-looking visible images. The approach generalizes well to complex scenes such as remote sensing and can also be applied to single-band infrared image colorization, improving the applicability of infrared image colorization.

\subsection{Vision Transformer and Wavelet Transform}

\begin{figure*}[htbp]
    \centering
    \includegraphics[scale=0.265, trim=0 0 0 0, clip]{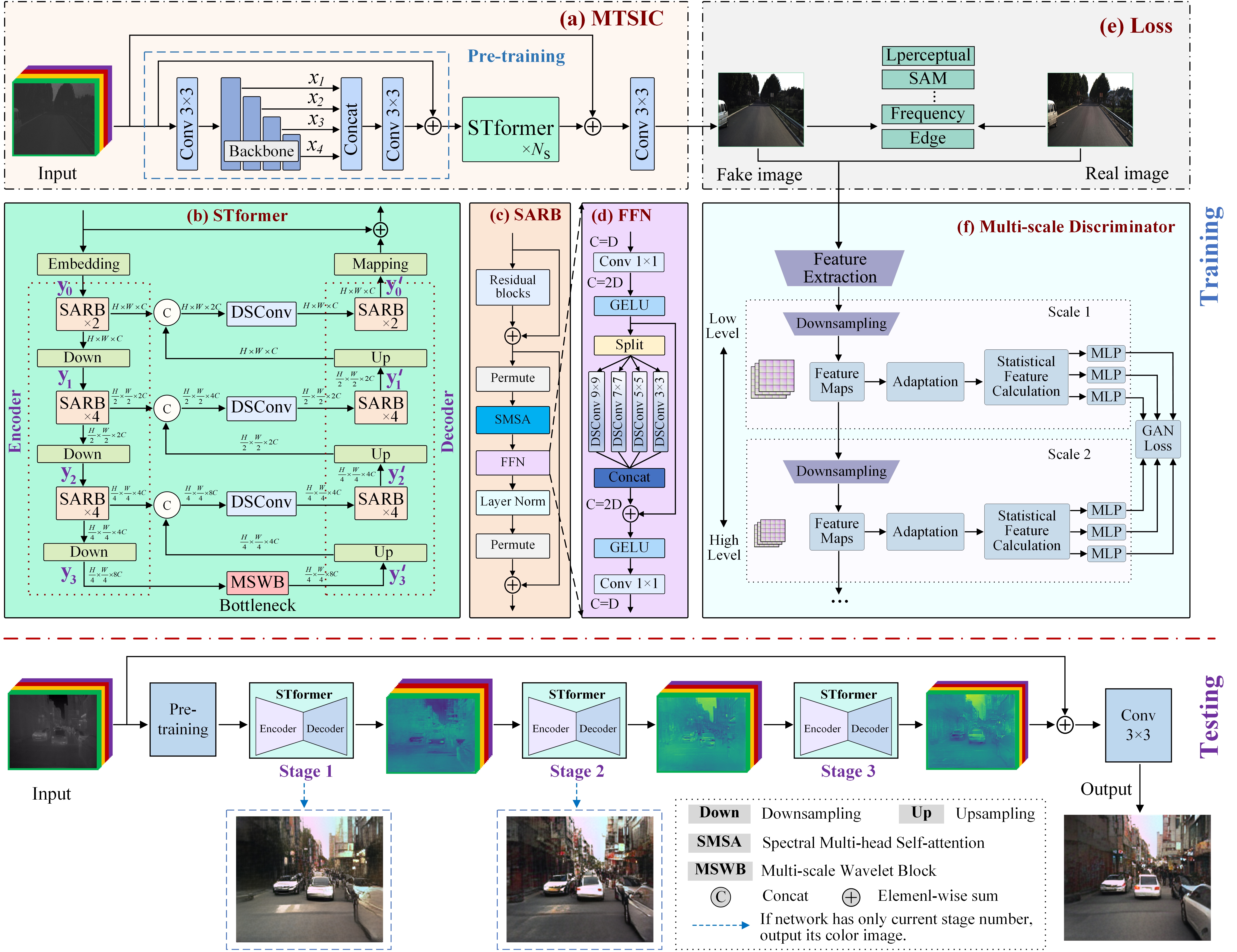}
    \caption{Overall architecture of the MTSIC.}
    \label{fig:Overall architecture}
\end{figure*}

Originally designed for machine translation, Transformers have been widely adopted in vision tasks like super-resolution \cite{[21]}, classification \cite{[40]}, and object detection \cite{[42]}. However, traditional Transformers focus on spatial modeling and are less effective for spectral images due to spatial sparsity and strong spectral correlations, making them ill-suited for handling spectral redundancy. To address this, we introduce Spectral Multi-head Self-Attention (SMSA) for infrared image colorization. SMSA computes self-attention along the spectral dimension, capturing both local and global spectral dependencies for improved representation. While GAN-based generators typically use ResNet or UNet variants, Transformer-based approaches for multi-band spectral image colorization remain largely unexplored.

In addition, frequency-based deep learning methods have gained interest. Most infrared colorization methods rely on spatial features and overlook frequency information. CNNs are biased toward low-frequency components \cite{[22],[26]}, prompting recent efforts to incorporate high-frequency representations—such as Fourier \cite{[50]} and Haar wavelet \cite{[51]} transforms—into vision tasks \cite{[26]}. Frequency-domain analysis complements spatial features, particularly in regions of infrared images with significant grayscale variations. Since wavelet transforms provide both spatial and frequency information, support multi-resolution analysis, and are computationally efficient \cite{[26]}. In this work, we integrate wavelet transforms into the Transformer framework to enhance multi-band infrared spectral image colorization.

\section {Proposed Method}

This section first provides an overview of the proposed MTSIC network architecture, followed by detailed introductions to the single-stage STformer, SARB, and the multi-scale wavelet block (MSWB). Finally, the design of the discriminator and the loss functions is described.

\subsection{Overall Architecture}

The overall architecture of MTSIC is shown in \autoref{fig:Overall architecture} (a). A pre-trained ConvNeXt \cite{[66]} is employed to efficiently extract spatial features, inspired by pixel-wise semantic segmentation to meet the fine-grained requirements of colorization. Given an input image $I_x \in \mathbb{R}^{L \times H \times W}$, where $L$ is the number of spectral channels, a $3 \times 3$ convolution is first applied to map the channels, followed by ConvNeXt to extract four hierarchical features:
$x_1 \in \mathbb{R}^{C \times H \times W}$,
$x_2 \in \mathbb{R}^{2C \times (H/2) \times (W/2)}$,
$x_3 \in \mathbb{R}^{4C \times (H/4) \times (W/4)}$,
$x_4 \in \mathbb{R}^{8C \times (H/8) \times (W/8)}$,
where $C = 96$. Features $x_2$, $x_3$, and $x_4$ are individually aligned via interpolation and then concatenated with the feature $x_1$. The concatenated features are subsequently fused through convolution, enhancing the representation of both local details and global structures.

The MTSIC comprises $N_s = 3$ cascaded STformer modules that reconstruct the corresponding daytime visible image from pre-trained spectral features. Local and global skip connections alleviate vanishing gradients and enhance feature reuse, improving training stability. The fusion of shallow and deep features further boosts convergence and representation.

\subsection{Single-stage Transformer-based Network}

\begin{figure*}[htbp]
    \centering
    \includegraphics[scale=0.49, trim=0 0 0 0, clip]{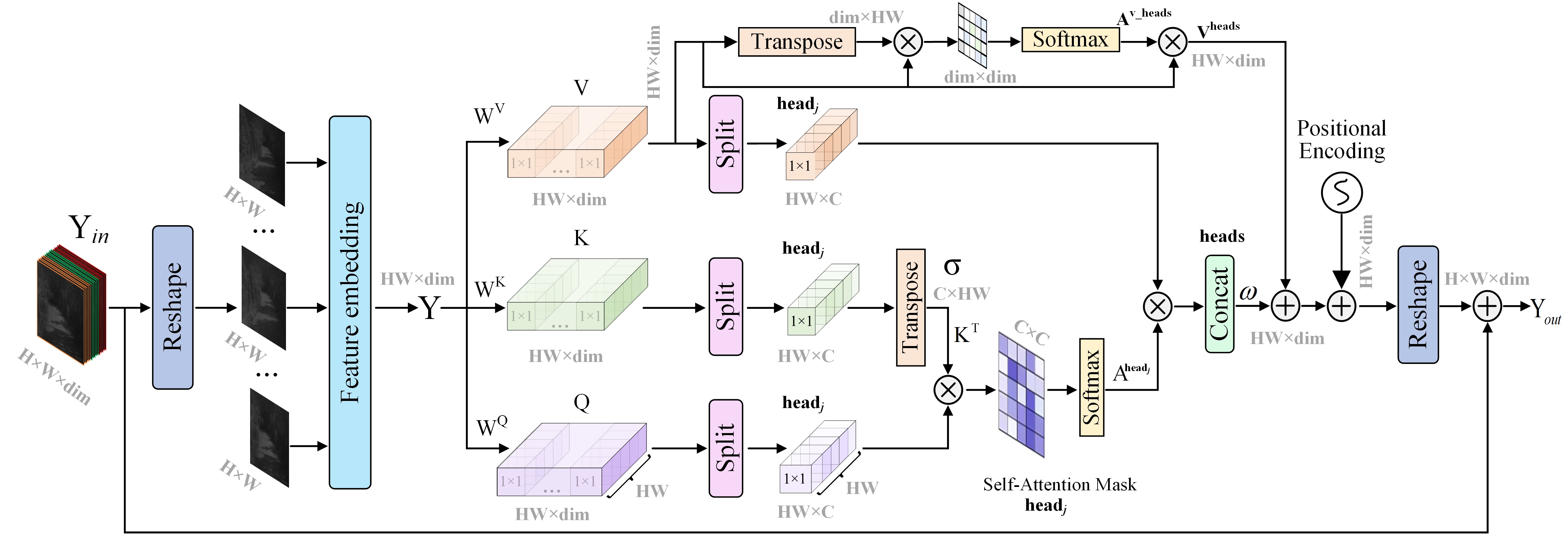}
    \caption{Spectral Self-Attention Mechanism.}
    \label{fig:SMSA}
\end{figure*}

\autoref{fig:Overall architecture} (b) illustrates the U-shaped STformer architecture, which consists of an encoder, decoder, and a bottleneck containing the MSWB module. Both the embedding and mapping layers use single-layer 3$\times$3 convolutions. In the encoder, features pass through the embedding layer, two SARB modules, two $4 \times 4$ downsampling convolutions (stride=4), and four additional SARB modules. The decoder mirrors this structure, using 2$\times$2 transposed convolutions (stride = 2) for upsampling. Features from the encoder and decoder are fused via channel-wise concatenation followed by a 1$\times$1 depthwise separable convolution (DSConv), enhancing multi-resolution representation and reducing computational complexity. This fusion strengthens the connection between local and global information, improving overall network performance.
 
\autoref{fig:Overall architecture} (c) illustrates the structure of the SARB, which consists of a residual block, a feed-forward network (FFN, as shown in \autoref{fig:Overall architecture} (d)), spectral multi-head self-attention, and two layer normalization operations.

\textbf{Spatial-Spectral Attention Residual Block (SARB):} The module fuses spatial-spectral information through multiple sub-units. Residual blocks first extract spatial features from the input \( \mathbf{y}_{in} \), improving spatial information modeling. A spectral multi-head self-attention (SMSA) mechanism then captures inter-band correlations to enhance spectral features. Finally, a feed-forward network (FFN) learns the mapping between spatial and spectral features using nonlinear activation and fully connected layers. Residual connections ensure efficient information flow, boosting expressiveness and mitigating overfitting.

Spectral Multi-head Self-attention (SMSA): Assuming the input $Y_{\text{\textit{in}}} \in \mathbb{R}^{H \times W \times \textit{dim}}$ is used as the input to the SMSA, it is first reshaped into a matrix $Y \in \mathbb{R}^{HW \times \textit{dim}}$. Then, $Y$ is linearly projected to obtain the \textbf{Query} matrix $Q \in \mathbb{R}^{HW \times \textit{dim}}$, \textbf{Key} matrix $K \in \mathbb{R}^{HW \times \textit{dim}}$, and \textbf{Value} matrix $V \in \mathbb{R}^{HW \times \textit{dim}}$, which are computed as follows:
\begin{equation}
\mathbf{Q}=\mathbf{Y} \mathbf{W}^{\mathbf{Q}}, \mathbf{K}=\mathbf{Y} \mathbf{W}^K, \mathbf{V}=\mathbf{Y} \mathbf{W}^{\mathbf{V}}
\end{equation}
The learnable projection weight matrices \(W^Q\), \(W^K\), and \(W^V\) are used for the Query, Key, and Value, respectively, forming the basis for spectral self-attention computation. For simplicity, biases are omitted. The model follows a U-shaped architecture with multiple SARB modules, where the feature dimension changes dynamically across layers. The number of heads \(N\) is determined by the feature dimension \(\textit{dim}\) and base dimension \(C\), with \(N = \textit{dim} \mathbin{//} C\). The feature dimension doubles in the encoder and halves in the decoder, ensuring their symmetry. \(Q\), \(K\), and \(V\) are split into \(N\) heads along the spectral channel: \(Q = [Q_1, \dots, Q_N]\), \(K = [K_1, \dots, K_N]\), and \(V = [V_1, \dots, V_N]\), each with dimension \(d_h = C\).

Note that \autoref{fig:SMSA} shows the case for $N=1$, with some details omitted for simplicity. Unlike conventional multi-head self-attention (MSA), the proposed SMSA treats each spectral representation as an independent token. Efficient mapping across bands is achieved through learnable query tokens and multi-head spectral self-attention. The self-attention computation for the $j$-th head (\textit{head}$_j$) is formulated as follows:
\begin{equation}
\mathbf{A}_j=\operatorname{softmax}\left(\sigma_j \mathbf{K}_j^{\mathrm{T}} \mathbf{Q}_j\right), \text { head }_j=\mathbf{V}_j \mathbf{A}_j,
\end{equation}
\noindent
where, $\mathbf{K}_j^{\mathrm{T}}$ denotes the transpose of $\mathbf{K}_j$. Due to spectral density variation with wavelength, a learnable parameter $\sigma_j \in \mathbb{R}^1$ is introduced to generate a new weight matrix via $\mathbf{K}j^{\mathrm{T}} \mathbf{Q}j$, which adjusts the self-attention weights $A_j$ for the $j$-th head. The outputs of the $N$ heads are then concatenated, reorganized, and linearly projected to restore the original dimensionality. However, this ignores spectral correlations between heads. To address this, a dual-attention mechanism is introduced. First, before multi-head splitting, following the aforementioned procedure, spectral self-attention $V_{\text{heads}}$ is computed from the matrix $\textbf{V}$, to capture correlations across the entire spectral range. Furthermore, $V_{\text{heads}}$ is fused with the attention outputs of individual heads to enhance spectral continuity. 
\begin{equation}
S M S A(Y)=\left(\text { Concat }_{j=1}^N\left(\text { head }_j\right)\right) W+V^{\text {heads }}+f_p(V)
\end{equation}
where, \( W \in \mathbb{R}^{C \times C} \) is a learnable weight matrix, and \( f_p(\cdot) \) represents the position encoding function, implemented using two depthwise separable 3×3 convolutions followed by GELU activation. Position encoding captures the spectral channel's positional information, as the spectral dimension is arranged by wavelength. By dynamically adjusting the number of heads, the model balances efficiency and capacity, effectively capturing spectral dependencies. The output from Equation (3) is reshaped and fused with shallow input features to enhance representation, producing the final output \( Y_{\text{\textit{out}}} \in \mathbb{R}^{H \times W \times C} \).

\begin{figure*}[htbp]
\centering
\includegraphics[scale=0.49, trim=0 0 0 0, clip]{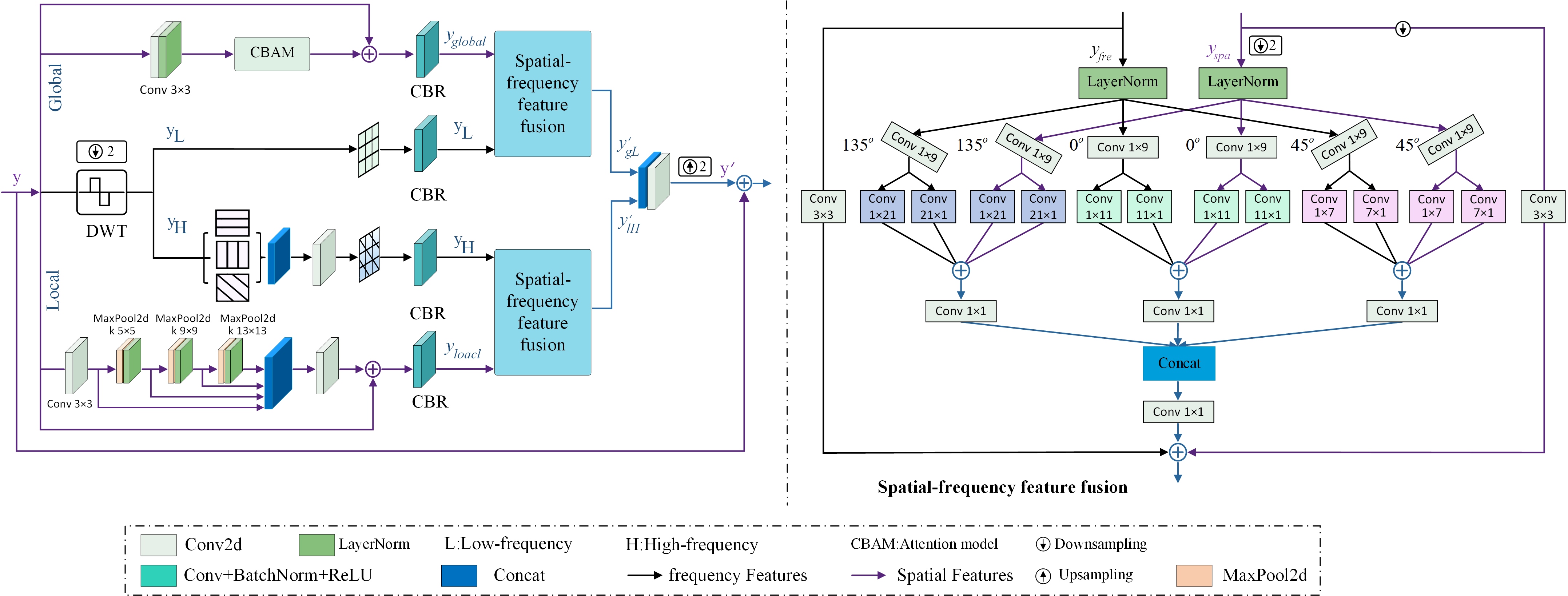}
    \caption{Multi-scale Wavelet Block.}
    \label{fig:MSWB_architecture}
\end{figure*}

\textbf{Multi-scale Wavelet Block (MSWB):} As depicted in \autoref{fig:MSWB_architecture}, the MSWB consists of three branches to enhance feature selection and representation. The global branch employs CBAM \cite{[54]} to enhance key feature identification through global spatial and channel recalibration, offering lower computational overhead compared to SMSA. The local branch employs consecutive multi-scale spatial pyramid pooling (5\(\times\)5, 9\(\times\)9, 13\(\times\)13) to extract features with varying receptive fields, while LayerNorm is incorporated to enhance the network’s nonlinear representation capability. Multi-scale information is fused through dimensional concatenation and convolution, effectively enriching detail representation. 
Since infrared images are discrete non-stationary signals containing multiple frequency components, the Haar wavelet transform is employed for multi-resolution decomposition to effectively capture local features and adapt to signal characteristics. In this work, the transform is used to map features into the frequency domain, decomposing them into a low-frequency component and three high-frequency components corresponding to horizontal, vertical, and diagonal directions, thereby enhancing edge and texture extraction. The wavelet branch converts spatial features into high-frequency features rich in local details and low-frequency features containing global information, providing more comprehensive frequency domain information. All branches utilize CBR units composed of convolution, Batch Normalization, and ReLU, where Batch Normalization improves training stability and ReLU enhances nonlinear representation capability. 
\begin{equation}
\left\{\begin{array}{l}
y_{\text {global }}=C B R\left(f_g(Y)\right) \\
y_{\text {local }}=C B R\left(f_m(Y)\right) \\
y_L, y_H=C B R\left(f_{w \downarrow 2}(Y)\right)
\end{array}\right.
\end{equation}
where $Y$ is the STformer encoder output, and $f_g(\cdot)$, $f_m(\cdot)$, and $f_w(\cdot)$ represent the global, local, and Haar wavelet transform branches, respectively. The features are denoted as $y_global$ (global), $y_{local}$ (multi-scale local), $y_L$ (low-frequency), and $y_H$ (high-frequency). The Haar wavelet transform downsamples by a factor of 2 ($\downarrow_2$). 

Subsequently, the spatial features and frequency-domain features are jointly fed into a spatial-frequency feature fusion module (SFFM), which aligns and integrates local high-frequency and global low-frequency information to obtain more discriminative hybrid features. Finally, the fused features are upsampled by a factor of 2 ($\uparrow2$), and a global skip connection integrates shallow features to further improve performance. 
\begin{equation}
\begin{gathered}
\left\{\begin{array}{l}
y_{g L}^{\prime}=f_{SFFM}\left(y_{g l o b a l}^{\prime}, y_L^{\prime}\right) \\
y_{l H}^{\prime}=f_{SFFM}\left(y_{l o b a l}^{\prime}, y_H^{\prime}\right)
\end{array}\right. \\
y^{\prime}=\operatorname{Conv}\left(\operatorname{Cat}\left[y_{g L}^{\prime}, y_{l H}^{\prime}\right]\right)_{\uparrow_2}+y
\end{gathered}
\end{equation}
where, \( f_{\text{SFFM}}(\cdot) \) denotes the spatial-frequency feature fusion module. 
\( y_{gL}' \) and \( y_{lH}' \) represent the global features fused with frequency-domain information 
and the multi-scale local features, respectively. 
The final output \( y' \) is obtained by fusing the shallow input features and the hybrid features via a global residual connection.

The MSWB adopts a simple structure, avoiding the complexity of STformer. By fusing spatial and frequency features, it expands the representation space while retaining frequency information, improving accuracy and robustness in category recognition. Next, we introduce its core component—the SFFM module for semantic fusion.

The SFFM employs large strip convolution kernels (\(1 \times k\) and \(k \times 1\)) for multi-scale feature extraction, which expands the receptive field and reduces parameters. To overcome the fixed-direction limitation, direction-aware strip convolutions~\cite{Directional_strip_convolution} (\(45^\circ \, \nearrow\), \(0^\circ \, \rightarrow\), \(135^\circ \, \searrow\)) are introduced, enabling flexible multi-directional feature extraction. Feature fusion is performed via dimensional concatenation and \(1 \times 1\) convolutions. Additionally, the SFFM adopts a mirrored structure to fuse features from two branches. After extracting features with strip convolutions, the convolution results along the same direction are cross-fused between two branches. On this basis, local features extracted by 3×3 square convolutions are added to complement the linear features from the strip convolutions, further enhancing the feature representation capability.

\begin{equation}
\left\{
\begin{aligned}
y_{\text{SConv}_7} &= \operatorname{Conv}_1\Big( 
    \operatorname{SConv}_7\big( 
        \operatorname{SConv}_{\nearrow}^{1 \times 9}(
            \operatorname{LN}(y_{\text{spal}_{\downarrow 2}})
        )
    \big) \\
    &\quad +\ \operatorname{SConv}_7\big(
        \operatorname{SConv}_{\nearrow}^{1 \times 9}(
            \operatorname{LN}(y_{\text{fre}})
        )
    \big) 
\Big) \\
y_{\text{SConv}_{11}} &= \operatorname{Conv}_1\Big( 
    \operatorname{SConv}_{11}\big( 
        \operatorname{SConv}_{\rightarrow}^{1 \times 9}(
            \operatorname{LN}(y_{\text{spal}_{\downarrow 2}})
        )
    \big) \\
    &\quad +\ \operatorname{SConv}_{11}\big(
        \operatorname{SConv}_{\rightarrow}^{1 \times 9}(
            \operatorname{LN}(y_{\text{fre}})
        )
    \big)
\Big) \\
y_{\text{SConv}_{21}} &= \operatorname{Conv}_1\Big( 
    \operatorname{SConv}_{21}\big( 
        \operatorname{SConv}_{\searrow}^{1 \times 9}(
            \operatorname{LN}(y_{\text{spal}_{\downarrow 2}})
        )
    \big) \\
    &\quad +\ \operatorname{SConv}_{21}\big(
        \operatorname{SConv}_{\searrow}^{1 \times 9}(
            \operatorname{LN}(y_{\text{fre}})
        )
    \big)
\Big)
\end{aligned}
\right.
\label{eq:ySConv}
\end{equation}

\begin{flalign}
&\mathrm{y}_{\text{mix}}^{\prime} = \operatorname{Conv}_1\Big(
    \operatorname{Cat}\big[
        y_{\text{SConv}_7},\ 
        y_{\text{SConv}_{11}},\ 
        y_{\text{SConv}_{21}}
    \big]
\Big) &\\
&\mathrm{y}_{\text{mix}}^{\prime} = \mathrm{y}_{\text{mix}}^{\prime} + \operatorname{Conv}_3(y_{\text{spal}_{\downarrow 2}}) + \operatorname{Conv}_3(y_{\text{fre}}) &
\end{flalign}
where, $y_{\mathrm{spa} \downarrow 2}$ denotes the spatial-domain input features after downsampling. To aggregate features from different branches, the spatial-scale features are downsampled at the same rate as in the wavelet transform, denoted as $\downarrow 2$. $y_{\mathrm{fre}}$ represents the input features in the frequency domain. 
$\operatorname{\textit{Conv}}_k$ denotes a $k \times k$ convolution, $\operatorname{\textit{SConv}}_k$ refers to asymmetric convolutions with kernel sizes of $1 \times k$ and $k \times 1$, $SConv_{\nearrow \rightarrow \searrow}^{1 \times 9}$ refers to directional strip convolutions applied along the directions (\(\nearrow\), \(\rightarrow\), \(\searrow\)), and $\operatorname{\textit{LN}}(\cdot)$ represents layer normalization. 
$y_{\text{SConv}_7}$, $y_{\text{SConv}_{11}}$, and $y_{\text{SConv}_{21}}$ denote the output features of directional convolutions (\(\nearrow\), \(\rightarrow\), \(\searrow\)) with kernel sizes of $1 \times k$ and $k \times 1$. $y_{\operatorname{mix}}^{\prime}$ denotes the output features of the multi-scale module, corresponding to $y_{gL}^{\prime}$ and $y_{lH}^{\prime}$ in Equation~(5), respectively.

\subsection{Discriminator}

The SPatchGAN \cite{[55]} discriminator, depicted in \autoref{fig:Overall architecture} (f), improves discrimination via multi-scale feature extraction and statistical computation. It takes an image $x$ (real or generated) and outputs $\mathcal{D}_{m,n}(x)$, where $m$ is the scale index and $n$ is the statistical feature index. The structure consists of an initial feature extraction block and $M$ scale modules, each containing a downsampling block, an adaptation block, a statistical feature computation block, and $N$ MLPs. A simplified version with two scale modules is provided in \autoref{fig:Overall architecture} (f). Feature maps are generated, reused at higher scales, and adjusted by the adaptation block. After extraction, image features go through downsampling, adaptation, statistical computation, and MLPs to produce scalar outputs.  

Unlike traditional PatchGAN \cite{[29]}, which shares convolutional weights, SPatchGAN uses global statistical features for more stable and accurate discrimination, overcoming local receptive field limitations. Each feature is processed by an independent MLP, improving accuracy. Statistical matching is more reliable at lower scales, though harder at higher scales, where it approximates the input distribution. SPatchGAN aims to match the distribution of statistical features, not the input image distribution. We adopt key parameter settings from the SPatchGAN discriminator.

\subsection{Loss Function}

The loss function measures the discrepancy between predictions and ground truth, guiding optimization. In infrared image colorization, a well-designed composite loss function improves translation quality. Based on this, a composite loss function for spectral infrared image colorization is developed:

\textbf{cGAN Loss:} The traditional GAN model maps random noise to a specific data distribution, but the process is unstable and the output is uncontrollable. Conditional GAN (cGAN) introduces constraints on both the generator and discriminator, providing additional guidance during learning \cite{[29]}. The cGAN loss function is as follows:
\begin{equation}
\begin{aligned}
\mathcal{L}_{\mathrm{cGAN}}(G, D) = & \mathbb{E}_{\mathrm{I}_x, \mathrm{I}_y}\left[\log D\left(\mathrm{I}_x, \mathrm{I}_y\right)\right] \\
& + \mathbb{E}_{\mathrm{I}_x, z}\left[\log \left(1-D\left(\mathrm{I}_x, G\left(\mathrm{I}_x\right)\right)\right)\right]
\end{aligned}
\end{equation}
where, $\mathrm{I}_x$ and $\mathrm{I}_y$ denote the input spectral infrared image and its corresponding ground truth, respectively. The generator \textit{G} seeks to minimize the objective function, while the discriminator \textit{D} aims to maximize it.

\textbf{Pixel Loss:} In image reconstruction tasks, $\mathcal{L}_1$ loss is commonly used as it penalizes pixel-level errors, aiding convergence. In this work, $\mathcal{L}_1$ spatial pixel loss is employed to improve image detail recovery during colorization.
\begin{equation}
\mathcal{L}_{\text {pixel }}(\Theta)=\frac{1}{\mathrm{~N}} \sum_{i=1}^N\left\|\left(I_{G T}\right)_i-G\left(I_x\right)_i\right\|_1
\end{equation}
where, $\Theta$ represents the learnable network parameters, $I_{G T}$ and $G\left(I_x\right)$ represents the Ground truth and the generated colorized image, respectively.

To minimize the chroma and luminance discrepancies between the generated colorized images and the ground truth, we introduce SAM loss $\mathcal{L}_{\text{SAM}}$ \cite{[16]}, which measures channel similarity by computing the angle between spectral vectors. Based on pixel-space constraints, we incorporate a frequency-domain loss $\mathcal{L}_{\text{fft}}$ \cite{fre_loss} to optimize low- and high-frequency components in both spatial and frequency domains, enhancing structure and details. Edge Loss $\mathcal{L}_{\text{edge}}$ \cite{edge_loss} is introduced to penalize reconstruction errors in edge regions, improving edge detail recovery.

\textbf{Objective Function:} The combined loss model of $\mathcal{L}_{\mathrm{cGAN}}$, $\mathcal{L}_{\text{pixel}}$, $\mathcal{L}_{\text{SAM}}$, $\mathcal{L}_{\text{fft}}$, and $\mathcal{L}_{\text{edge}}$ is adopted as the initial configuration of the proposed method. Building upon this setup and inspired by the high-performing loss functions in infrared image colorization tasks reported by Kuang et al.~\cite{[16]}, perceptual loss $\mathcal{L}_{\text{per}}$, total variation loss $\mathcal{L}_{\text{tv}}$, and structural similarity loss $\mathcal{L}_{\text{SSIM}}$ are incorporated as additional terms to enhance the realism and quality of the generated images. The objective function of the proposed method is defined as follows:
\begin{equation}
\begin{aligned}
\mathcal{L}_{\text{total}} =\ & \lambda_{\text{cGAN}} \mathcal{L}_{\text{cGAN}} + \lambda_{\text{pix}} \mathcal{L}_{\text{pixel}} 
+ \lambda_{\text{SAM}} \mathcal{L}_{\text{SAM}} + \lambda_{\text{fft}} \mathcal{L}_{\text{fft}} \\
& + \lambda_{\text{edge}} \mathcal{L}_{\text{edge}} + \lambda_{\text{per}} \mathcal{L}_{\text{per}} 
+ \lambda_{\text{tv}} \mathcal{L}_{\text{tv}} + \lambda_{\text{ssim}} \mathcal{L}_{\text{SSIM}}
\end{aligned}
\end{equation}
where $\lambda$ denotes the hyperparameter controlling the weighting of each loss function. Empirically, the values are set as $\lambda_{\text{cGAN}} = 1$, $\lambda_{\text{pix}} = 50$, $\lambda_{\text{SAM}} = 0.1$, $\lambda_{\text{edge}} = 0.5$, and $\lambda_{\text{per}} = \lambda_{\text{tv}} = \lambda_{\text{ssim}} = \lambda_{\text{fft}} = 1$ to ensure that all loss terms remain within the same order of magnitude. This hybrid loss function is designed to optimize network performance through spatial, spectral, and frequency domain reconstruction accuracy.

\section{Experiments}

This section introduces the infrared spectral dataset, evaluation metrics, and implementation details. It then compares the proposed method with existing approaches and conducts ablation studies to assess the contributions of individual modules and loss functions.

\subsection{Datasets}

(a) The KAIST dataset\footnote{\url{https://github.com/SoonminHwang/rgbt-ped-detection}} is a multispectral pedestrian detection dataset containing 95,000 paired thermal and RGB images from day and night scenes in campus, road, and urban environments \cite{KAIST_Dataset}. Due to low contrast and noise, thermal images pose detection challenges, which can be mitigated by colorization to enhance detail, contrast, and detection accuracy.

(b) The HSI ROAD dataset\footnote{\url{NUST-Machine-Intelligence-Laboratory/hsi_road}} is a real-world hyperspectral road segmentation dataset with 28 spectral bands, consisting of 3,799 pairs of RGB and near-infrared images \cite{HSIROAD_Dataset}. Unlike typical urban RGB datasets, it includes diverse road surfaces—such as asphalt, concrete, soil, and sand—under rural and natural conditions, enhancing diversity and representativeness for segmentation tasks.

(c) The IHSR dataset is a self-collected long-wave infrared hyperspectral remote sensing dataset captured in Hengdian Town, Zhejiang, China, covering diverse land-cover types including mountains, urban areas, rivers, parks, highways, and farmland. It offers 110 spectral bands spanning 8.0--11.3\,$\mu$m at a 1.0\,m spatial resolution, with dimensions of $443 \times 17189 \times 110$. Corresponding labeled RGB images are $443 \times 17189 \times 3$.

\subsection{Implementation details}

Experiments were conducted on the KAIST, HSI ROAD, and IHSR datasets. For KAIST, nighttime campus, road, and urban scenes were used, with thermal-RGB pairs resized to $500 \times 400$ and center-cropped to $360 \times 288$. HSI ROAD images, originally $192 \times 384$, were resized to $512 \times 450$, each containing RGB and near-infrared hyperspectral pairs. For IHSR, pixel-level registration in ENVI was followed by selecting 20 non-overlapping regions for testing and the rest for training. Due to limited spatial size, overlapping block sampling (stride 24) was used to extract $354 \times 354$ training patches.

All datasets were augmented with random cropping to $256 \times 256$, rotation, and scaling. Models were trained for 60 epochs with an initial learning rate of $1 \times 10^{-4}$, decaying linearly after epoch 30. A batch size of 1 and the Adam optimizer \cite{Adam} were used. Experiments were run on Ubuntu 18.04 with PyTorch 2.0.1 and an RTX 3090 GPU.

\subsection{Evaluation Metrics}

Due to the absence of corresponding daytime visible-light labeled images in the KAIST dataset, semantic segmentation and object detection are used to evaluate the texture and color naturalness of the reconstructed images. For the HSI ROAD and IHSR datasets, network performance is evaluated using four image quality metrics (PQI): PSNR, SSIM \cite{SSIM}, NIQE \cite{[60]}, and UIQI \cite{[61]}. NIQE measures image naturalness by quantifying deviations from high-quality natural image models, while UIQI evaluates luminance, contrast, and structural similarity to the ground truth. PSNR and SSIM focus on content quality, with ideal values of $+\infty$ and $1$, respectively. NIQE and UIQI assess color quality, with ideal values of $0$ and $1$.

In addition, a no-reference metric called Colorful is introduced, with an ideal value of 1, to subjectively assess the color richness of generated images. This indicator captures critical details affecting the visual experience and is based on the evaluation model proposed in \cite{color_metric}. To further evaluate the color distribution differences between the generated images and ground truth, this paper proposes a novel metric named ColorJSD. Unlike pixel-level metrics such as NIQE and UIQI, ColorJSD focuses on the similarity of overall color distributions based on probabilistic divergence. The expression of ColorJSD is defined as follows:
\begin{equation}
\operatorname{ColorJSD}(P \| Q)=\frac{1}{2} D_{K L}(P \| M)+\frac{1}{2} D_{K L}(Q \| M)
\end{equation}
where, \( P \) and \( Q \) represent the color distribution histograms of the generated and ground truth images, respectively, and $M=\frac{P+Q}{2}$ denotes their average distribution. $D_{K L}(P \| Q)$ refers to the Kullback–Leibler (KL) divergence, calculated as follows:
\begin{equation}
D_{K L}(P \| Q)=\sum_i P(i) \log \left(\frac{P(i)}{Q(i)}\right)
\end{equation}
The ideal value of ColorJSD is 0, indicating that the color distribution of the generated image closely matches that of the ground truth.

\subsection{Comparisons With the State-of-the-Art Methods}

To validate the superiority of our method, eight representative infrared image colorization approaches were compared, including pix2pix~\cite{[29]}, TICC-GAN~\cite{[16]}, ToDayGAN~\cite{[62]}, FRAGAN\_P~\cite{[23]}, LKAT-GAN~\cite{[13]}, DDGAN~\cite{[18]}, MUGAN~\cite{MUGAN}, and MornGAN~\cite{[31]}. Among them, pix2pix and ToDayGAN are general image-to-image translation frameworks, commonly applied to infrared colorization tasks. The others are specifically designed for single-band infrared images. We trained and evaluated all models using their official PyTorch implementations and original parameter settings on the same dataset.

\textbf{(1) Experimental Results on the KAIST Multispectral Dataset}

\begin{table}[htbp]
  \centering
\setlength{\tabcolsep}{3.0pt} 
  \caption{The average semantic segmentation performance (IoU) of images generated by different colorization methods on 500 test images from the KAIST dataset is presented in the table, with red indicating the best performance and blue indicating the second-best performance.}
    \begin{tabular}{@{}ccccccc@{}}
    \toprule
    \textbf{Comparison} & \textbf{Road} & \textbf{Building} & \textbf{Sky} & \textbf{Person} & \textbf{Car} & \textbf{mIoU} \\
    \midrule
    Input NTIR image   & 70.5 & 20.2 & 8.7  & 8.0  & 6.9  & 22.86 \\
    Reference NVC image & \textcolor[rgb]{ .753,  0,  0}{82.1} & \textcolor[rgb]{ .753,  0,  0}{42.0} & 22.4 & \textcolor[rgb]{ .176,  .329,  .627}{18.5} & \textcolor[rgb]{ .176,  .329,  .627}{18.9} & \textcolor[rgb]{ .176,  .329,  .627}{36.78} \\
    pix2pix \cite{[29]} & 75.2 & 19.7 & 10.3 & 9.3  & 12.8 & 25.46 \\
    TICC-GAN \cite{[16]} & 80.0 & 28.7 & 16.1 & 13.2 & 15.7 & 30.74 \\
    TodayGAN \cite{[62]} & 76.9 & 22.1 & 13.4 & 10.8 & 10.5 & 26.74 \\
    DDGAN \cite{[18]} & 75.5 & 24.2 & 11.5 & 8.7  & 9.7  & 25.92 \\
    FRAGAN\_P \cite{[23]} & 79.7 & 29.9 & 15.8 & 11.3 & 13.1 & 29.96 \\
    LKAT-GAN \cite{[13]} & 81.2 & 32.4 & 17.1 & 14.6 & 17.1 & 32.48 \\
    MUGAN \cite{MUGAN} &  80.3 & 26.9 & 14.5 & 12.0 & 14.5 & 29.64 \\
    MornGAN \cite{[31]} & 70.7 & 20.2 & \textcolor[rgb]{ .176,  .329,  .627}{30.2} & 15.9 & 15.0 & 30.40 \\
    Ours  & \textcolor[rgb]{ .176,  .329,  .627}{81.5} & \textcolor[rgb]{ .176,  .329,  .627}{38.2} & \textcolor[rgb]{ .753,  0,  0}{30.4} & \textcolor[rgb]{ .753,  0,  0}{18.9} & \textcolor[rgb]{ .753,  0,  0}{19.2} & \textcolor[rgb]{ .753,  0,  0}{37.64} \\
    \bottomrule
    \end{tabular}%
  \label{tab:KAIST_Segmention}%
\end{table}%

\begin{table}[htbp]
  \centering
  \setlength{\tabcolsep}{2pt}
  \caption{Comparison of the average pedestrian detection performance of different colorization methods on 6112 test images from the KAIST dataset, computed at an IoU of 0.50. The best performance is highlighted in red, and the second-best in blue for each column.}
  \begin{tabular}{ccccc}
    \toprule
    \textbf{Comparison} & \textbf{Precision $\uparrow$} & \textbf{Recall $\uparrow$} & \textbf{mAP50 $\uparrow$} & \textbf{Colorful $\uparrow$} \\
    \midrule
    Input NTIR image & 63.6 & 39.5 & \textcolor[rgb]{0,.439,.753}{40.6} & 0.01 \\
    Reference NVC image & \textcolor[rgb]{0,.439,.753}{63.9} & 39.4 & 40.0 & 0.29 \\
    pix2pix \cite{[29]} & 12.4 & 16.2 & 5.9 & \textcolor[rgb]{0,.439,.753}{0.49} \\
    TICC-GAN \cite{[16]} & 40.6 & 25.3 & 22.7 & 0.31 \\
    TodayGAN \cite{[62]} & 31.9 & 21.9 & 16.1 & 0.34 \\
    DDGAN \cite{[18]} & 19.4 & 13.1 & 8.1 & 0.16 \\
    FRAGAN\_P \cite{[23]} & 39.9 & 26.1 & 23.1 & 0.46 \\
    LKAT-GAN \cite{[13]} & 44.8 & 31.6 & 28.8 & 0.36 \\
    MUGAN \cite{MUGAN} & 32.5 & 20.4 & 16.5 & 0.18 \\
    MornGAN \cite{[31]} & 48.1 & \textcolor[rgb]{.753,0,0}{42.1} & 37.7 & 0.43 \\
    Ours & \textcolor[rgb]{.753,0,0}{64.2} & \textcolor[rgb]{0,.439,.753}{41.3} & \textcolor[rgb]{.753,0,0}{40.8} & \textcolor[rgb]{.753,0,0}{0.54} \\
    \bottomrule
  \end{tabular}%
  \label{tab:KAIST_Objection}%
\end{table}%

\begin{figure*}[htbp]
    \centering
    \includegraphics[scale=0.17, trim=0 0 0 0, clip]{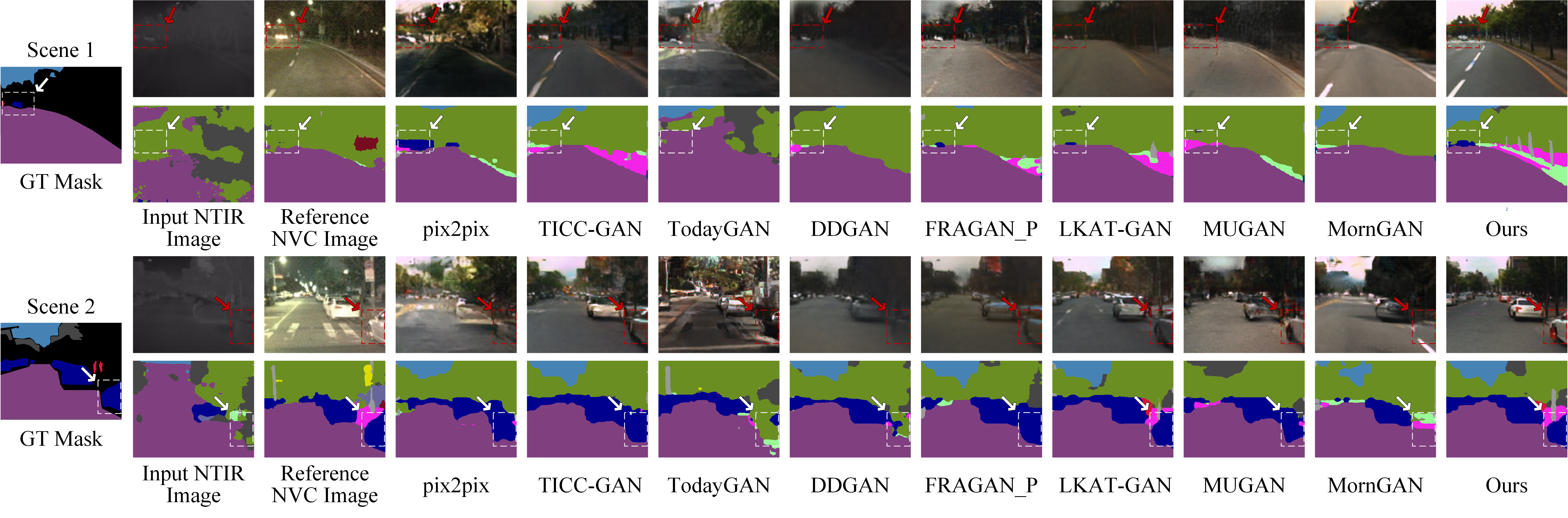}
    \caption{Visual comparison of colorization (top row) and segmentation (bottom row) performance of different methods on the KAIST dataset. The highlighted regions within the white dotted boxes in the segmentation results are of particular interest.}
    \label{fig:KAIST_Segmention}
\end{figure*}

\begin{figure*}[htbp]
    \centering
    \includegraphics[scale=0.195, trim=0 0 0 0, clip]{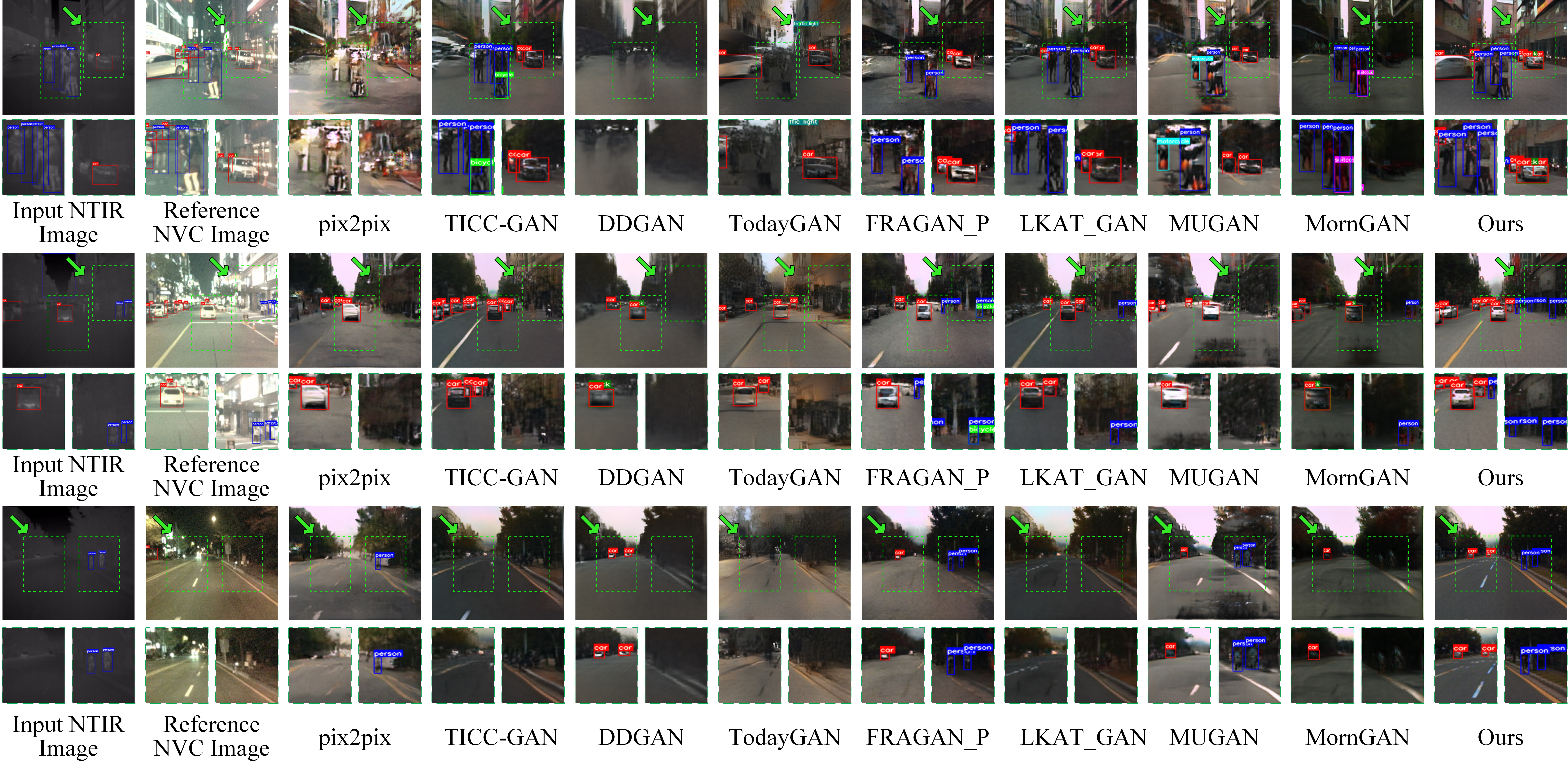}
    \caption{A visual comparison of pedestrian detection results using the YOLOv7 model \cite{[19]} on the KAIST dataset. Blue bounding boxes indicate “Person,” while red bounding boxes denote “Car.”}
    \label{fig:KAIST_Objection}
\end{figure*}

In the KAIST dataset test, we convert nighttime thermal infrared (NTIR) images into daytime RGB images. Lacking pixel-level aligned RGB labels, we use the Segformer model~\cite{SegFormer}, trained on the Cityscape dataset~\cite{[64]}, for semantic segmentation to assess the feature quality and semantic consistency of the converted images. As shown in \autoref{tab:KAIST_Segmention} and \autoref{fig:KAIST_Segmention}, our method significantly outperforms others in segmenting small targets (e.g., pedestrians, vehicles), even slightly surpassing the reference nighttime visible images (NVC). This advantage is especially clear in the ``sky'' category, where low illumination hampers segmentation in NTIR images.

Traditional methods (e.g., pix2pix~\cite{[29]}, TodayGAN~\cite{[62]}, DDGAN~\cite{[18]}, MUGAN~\cite{MUGAN}) struggle with NTIR’s low contrast and complex structures, resulting in poor segmentation of small targets. While MornGAN~\cite{[31]} performs relatively well, it still misrepresents ``building'' and ``vehicle'' categories. In contrast, TICC-GAN~\cite{[16]}, FRAGAN\_P~\cite{[23]}, LKAT-GAN~\cite{[13]}, and our method achieve better reconstruction and segmentation, with our approach excelling in both small-target detail and color realism. For example, in Scene~1, our model accurately reconstructs small vehicles and distinguishes ``tree trunks'' and service lanes. It achieves the highest mIoU, surpassing NVC and exceeding the next-best method, LKAT-GAN~\cite{[13]}, by 15.9\%. This gain is due to our multi-stage Transformer combined with wavelet transform, which enhances feature extraction and reconstruction, leading to improved segmentation accuracy.

To evaluate the realism of object features in generated images, we use YOLOv7~\cite{[19]}, trained on the MS COCO dataset~\cite{[20]}, for object detection. Results are presented in \autoref{tab:KAIST_Objection} and \autoref{fig:KAIST_Objection}. Methods like Pix2pix~\cite{[29]}, TodayGAN~\cite{[62]}, DDGAN~\cite{[18]}, and MUGAN~\cite{MUGAN} struggle with complex structures, often producing blurred building edges due to semantic confusion. In contrast, TICC-GAN~\cite{[16]}, FRAGAN\_P~\cite{[23]}, LKAT-GAN~\cite{[13]}, and MornGAN~\cite{[31]} improve edge clarity through advanced loss functions and generator designs.

As illustrated in the first two subfigures of \autoref{fig:KAIST_Objection}, most methods fail to reconstruct small targets (e.g., vehicles, pedestrians) accurately. Our method, however, preserves fine details, sharp edges, and realistic colors. Under low-light conditions, YOLOv7 detects only some pedestrians in NVC images, while our method captures full crowd poses and reduces missed detections. Although our recall is slightly lower than MornGAN’s, we achieve higher precision, leading to an 8.2\% gain in mean Average Precision (mAP). This confirms our model’s strengths in feature extraction, detail preservation, and colorization.

Additionally, as shown in \autoref{tab:KAIST_Objection}, although the input infrared images exhibit low Noise, their Colorful is relatively weak. After colorization, this metric is significantly improved, enhancing color representation and subjective perception, which demonstrates the necessity of the colorization process.

\textbf{(2) Experimental Results on the HSI ROAD and IHSR Hyperspectral Datasets}

\begin{figure*}[htbp]
    \centering
    \includegraphics[scale=0.20, trim=0 0 0 0, clip]{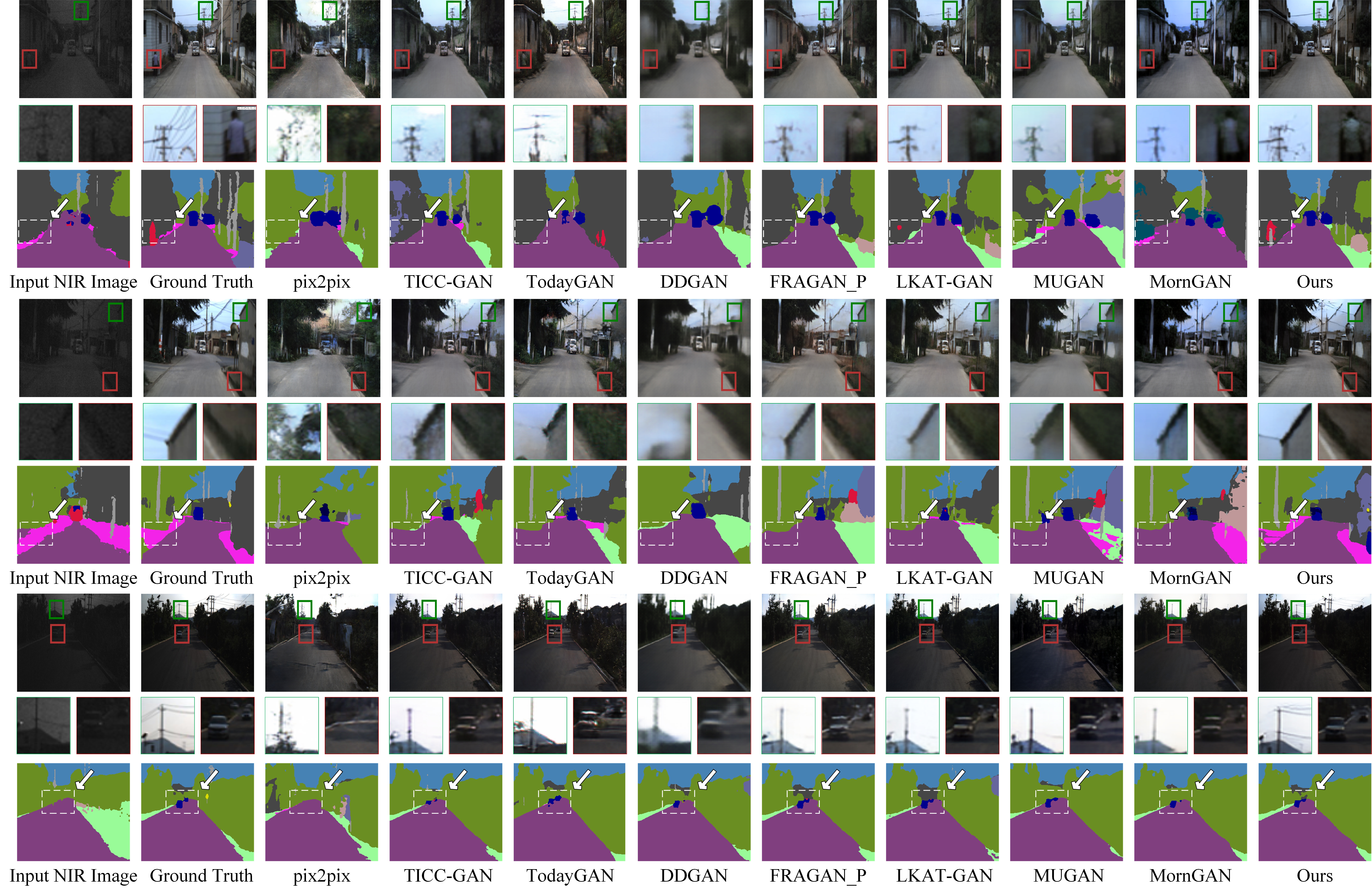}
    \caption{A comparative visualization of colorization (top panel) and segmentation (bottom panel) performance across different methods is presented for the HSI ROAD dataset. Particular attention should be directed to the regions demarcated by white dotted boxes in the segmentation results.}
    \label{fig:HSI_Road}
\end{figure*}

\begin{table*}[htbp]
  \centering
  \setlength{\tabcolsep}{3.8pt}
  \caption{Quantitative comparison on the HSI ROAD and IHSR datasets. "↑" indicates that higher values represent better performance, while "↓" indicates that lower values are preferred. The best and second-best results are highlighted in red and blue, respectively.}
  \label{tab:HSI_ROAD_IHSR}
  \begin{tabular}{c|cccccc|cccccc}
    \toprule
    \multicolumn{1}{c|}{} & \multicolumn{6}{c|}{\textbf{HSI ROAD dataset}} & \multicolumn{6}{c}{\textbf{IHSR dataset}} \\
    \midrule
    \textbf{Methods} & \textbf{PSNR↑} & \textbf{SSIM↑} & \textbf{NIQE↓} & \textbf{UIQI↑} & \textbf{Colorful↑} & \textbf{ColorJSD↓} & \textbf{PSNR↑} & \textbf{SSIM↑} & \textbf{NIQE↓} & \textbf{UIQI↑} & \textbf{Colorful↑} & \textbf{ColorJSD↓} \\
    \midrule
    pix2pix \cite{[29]}  & 16.93 & 0.51 & 4.25 & 0.83 & \textcolor[rgb]{0.753,0,0}{0.31} & 0.09 & 17.94 & 0.40 & 8.32 & 0.65 & \textcolor[rgb]{0.176,0.329,0.627}{0.59} & 0.12 \\
    TICC-GAN \cite{[16]} & 21.39 & 0.72 & 4.25 & 0.93 & 0.21 & 0.10 & 18.09 & 0.42 & 8.27 & 0.60 & 0.15 & 0.24 \\
    TodayGAN \cite{[62]} & 16.62 & 0.52 & \textcolor[rgb]{0.176,0.329,0.627}{3.87} & 0.81 & 0.11 & 0.37 & 12.41 & 0.17 & 8.37 & 0.30 & 0.09 & 0.19 \\
    DDGAN \cite{[18]}    & 15.36 & 0.54 & 4.25 & 0.74 & 0.10 & 0.10 & 21.07 & 0.60 & 8.25 & \textcolor[rgb]{0.176,0.329,0.627}{0.83} & 0.28 & \textcolor[rgb]{0.176,0.329,0.627}{0.11} \\
    FRAGAN\_P \cite{[23]} & 21.53 & 0.72 & 4.23 & \textcolor[rgb]{0.176,0.329,0.627}{0.94} & 0.17 & \textcolor[rgb]{0.176,0.329,0.627}{0.07} & \textcolor[rgb]{0.176,0.329,0.627}{21.20} & \textcolor[rgb]{0.176,0.329,0.627}{0.68} & \textcolor[rgb]{0.176,0.329,0.627}{8.24} & \textcolor[rgb]{0.176,0.329,0.627}{0.83} & 0.44 & 0.18 \\
    LKAT-GAN \cite{[13]} & 21.69 & \textcolor[rgb]{0.176,0.329,0.627}{0.74} & 4.23 & \textcolor[rgb]{0.176,0.329,0.627}{0.94} & 0.20 & 0.08 & 19.98 & 0.55 & 8.27 & 0.76 & 0.22 & 0.13 \\
    MUGAN \cite{MUGAN}   & \textcolor[rgb]{0.176,0.329,0.627}{21.70} & 0.73 & 4.23 & \textcolor[rgb]{0.176,0.329,0.627}{0.94} & \textcolor[rgb]{0.176,0.329,0.627}{0.26} & \textcolor[rgb]{0.176,0.329,0.627}{0.07} & 20.16 & 0.56 & 8.27 & 0.79 & 0.56 & 0.17 \\
    MornGAN \cite{[31]}  & 20.69 & 0.61 & 4.25 & 0.93 & 0.11 & 0.16 & 18.06 & 0.42 & 8.29 & 0.60 & 0.10 & 0.23 \\
    Ours                & \textcolor[rgb]{0.753,0,0}{22.54} & \textcolor[rgb]{0.753,0,0}{0.77} & \textcolor[rgb]{0.753,0,0}{3.32} & \textcolor[rgb]{0.753,0,0}{0.95} & 0.24 & \textcolor[rgb]{0.753,0,0}{0.05} & \textcolor[rgb]{0.753,0,0}{23.87} & \textcolor[rgb]{0.753,0,0}{0.81} & \textcolor[rgb]{0.753,0,0}{8.21} & \textcolor[rgb]{0.753,0,0}{0.94} & \textcolor[rgb]{0.753,0,0}{0.62} & \textcolor[rgb]{0.753,0,0}{0.07} \\
    \bottomrule
  \end{tabular}
\end{table*}

\begin{figure*}[htbp]
    \centering
    \includegraphics[scale=0.26, trim=0 0 0 0, clip]{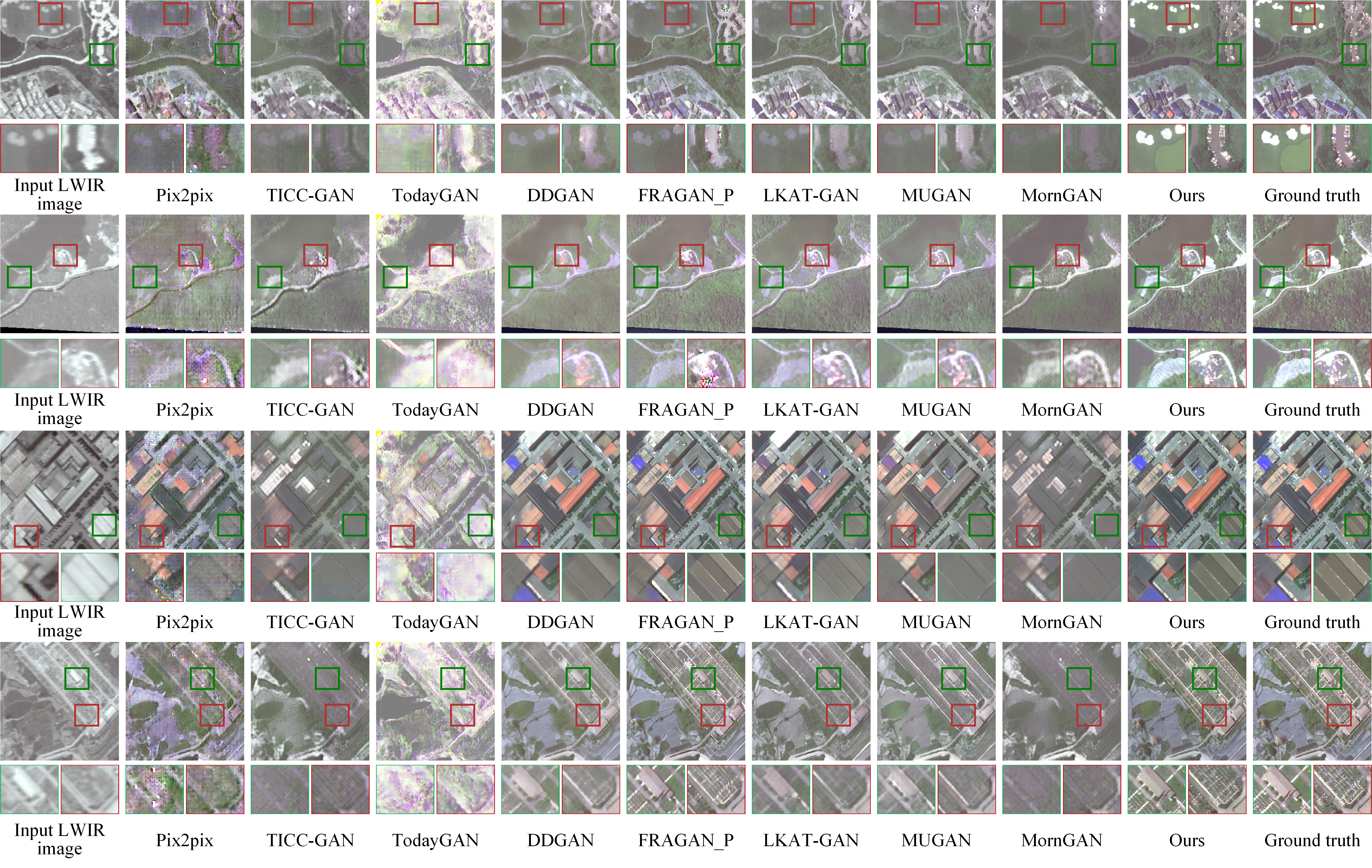}
    \caption{Visualization comparison of partial images from the Long-Wave Infrared Hyperspectral Remote Sensing Dataset (IHSR).}
    \label{fig:IHSI_Remote_sensing}
\end{figure*}

Compared to the KAIST dataset, the HSI ROAD and IHSR hyperspectral datasets enable better performance in image colorization. HSI ROAD's 28 spectral bands provide richer textures, helping most methods better reconstruct small object contours like vehicles, pedestrians, and buildings. As illustrated in \autoref{fig:HSI_Road}, Pix2pix~\cite{[29]}, TodayGAN~\cite{[62]}, and DDGAN~\cite{[18]} suffer from geometric distortions and missing details (e.g., Pix2pix fails to restore vehicle outlines; DDGAN omits building edges). While other methods retain global structure, they lack detail fidelity. In contrast, our method best preserves fine details such as utility poles, wires, and building edges. Segformer-based ~\cite{SegFormer} semantic segmentation further validates this: segmentation results from our generated images closely match the ground truth and outperform those from raw infrared inputs, reflecting higher reconstruction quality.

Quantitative results in \autoref{tab:HSI_ROAD_IHSR} confirm our advantage, with the highest PSNR, SSIM, UIQI, and lowest NIQE on HSI ROAD. Notably, our PSNR surpasses MUGAN~\cite{MUGAN} by 3.9\%, and NIQE is optimized by 16.6\% compared to TodayGAN~\cite{[62]}, indicating more natural and visually realistic outputs.

Compared to the HSI ROAD dataset, the IHSR dataset presents greater challenges due to its large-scale coverage, low spatial resolution (1.0 m), and the presence of 110 thermal infrared spectral bands. These characteristics demand careful modeling of spectral correlations during the colorization process. Despite the limited spatial resolution, the rich spectral information enhances semantic discrimination. As shown in \autoref{fig:IHSI_Remote_sensing}, our method achieves superior texture restoration and edge clarity compared to Pix2pix~\cite{[29]}, TICC-GAN~\cite{[16]}, and MUGAN~\cite{MUGAN}. It also outperforms FRAGAN\_P~\cite{[23]} and DDGAN~\cite{[18]} in detail sharpness and color fidelity, benefiting from targeted network optimizations. Although LKAT-GAN improves the generator architecture, it performs poorly on the IHSR dataset. This suggests that enhancing the generator alone, without adequately accounting for spectral correlations, leads to information loss during the infrared-to-color image cross-domain colorization process, thereby degrading reconstruction quality. As unsupervised methods, TodayGAN and MornGAN do not require labeled data; however, their texture and color restoration on the IHSR dataset are significantly inferior to supervised methods, highlighting the irreplaceable advantages of supervised learning. As shown in \autoref{tab:HSI_ROAD_IHSR}, our method achieves the best performance on the IHSR dataset, surpassing the second-best method (FRAGAN\_P) with improvements of 12.6\% in PSNR, 19.1\% in SSIM, and 13.3\% in UIQI. Furthermore, it demonstrates notable advantages in the {Colorful} and {ColorJSD} metrics, especially in {ColorJSD}. These results suggest that our approach more effectively extracts deep spatial features and fully leverages spectral correlations, enabling the reconstruction of more natural and vivid colors.

\subsection{Ablation study}

\begin{table}[htbp]
  \centering
  \setlength{\tabcolsep}{1.8pt} 
  \caption{Ablation study of stage number \(N_s\) and attention mechanism complexity}
    \begin{tabular}{ccccccc}
    \toprule
    \textbf{\makecell[c]{Comparison}} & \textbf{\makecell[c]{Variant}} & \textbf{\makecell[c]{Precision\\(\%)}} & \textbf{\makecell[c]{Recall\\(\%)}} & \textbf{\makecell[c]{mAP50\\(\%)}} & \textbf{\makecell[c]{\#Params\\(M)}} & \textbf{\makecell[c]{FLOPs\\(G)}} \\
    \midrule
    \multirow{4}{*}{\makecell[c]{Stage\\ number}} & Ns = 1 & 56.1  & 36.3  & 34.1  & 48.07 & 235.79 \\
          & Ns = 2 & 59.5  & 39.7  & 38.3  & 63.64 & 426.47 \\
          & Ns = 3 & 64.2  & 41.3  & 40.8  & 79.22 & 617.14 \\
          & Ns = 4  & \textbf{65.8} & \textbf{41.4} & \textbf{41.6} & 94.8  & 807.81 \\
    \midrule
    \multirow{4}{*}{\makecell[c]{Attention\\ Complexity}} & G-MSA \cite{GMSA} & 60.7  & 35.2  & 38.1  & 86.4  & 914.76 \\
          & W-MSA \cite{[67]} & 60.1  & 31.9  & 34.1  & 67.12 & 507.54 \\
          & SW-MSA \cite{[67]} & 57.3  & 38.7  & 36.7  & 62.17 & 501.43 \\
          & SMSA  & \textbf{64.2} & \textbf{41.3} & \textbf{40.8} & 79.22 & 617.14 \\
    \bottomrule
    \end{tabular}
  \label{tab:stage_number_attention_comparison}
\end{table}

\textbf{Stage Number Analysis:} A multi-stage learning strategy is introduced by cascading multiple STformers to achieve coarse-to-fine reconstruction optimization. The effect of the number of stages, $N_s$, on performance is analyzed (see \autoref{tab:stage_number_attention_comparison}). Results indicate that performance improves with higher $N_s$, but computational cost (parameters and FLOPs) increases linearly. While $N_s = 4$ offers slightly better performance than $N_s = 3$, it significantly increases computational cost. Thus, $N_s = 3$ is chosen for MTSIC, balancing performance and efficiency, and used as the default in subsequent experiments. As illustrated in \autoref{tab:stage_number_attention_comparison}, and compared with \autoref{tab:KAIST_Objection}, even with $N_s = 1$, the proposed method outperforms several single-band methods and has fewer parameters than high-performance models like LKAT-GAN (83.69M) and FRAGAN\_P (52.45M), highlighting the effectiveness and superiority of the proposed framework.

The Transformer-based STformer network includes two key modules: SARB and MSWB. For fair evaluation, only one variable was modified in each experiment while keeping other parameters consistent. Detailed analyses of SARB and MSWB follow.

\textbf{Complexity Analysis of SMSA:} According to \cite{GMSA}, G-MSA computes correlations by treating all spatial tokens as queries and keys, with the number of tokens given by \( n = H \times W \), resulting in high computational cost. In W-MSA \cite{[67]}, the feature map is divided into non-overlapping \( M \times M \) windows, and self-attention is computed within each window. This reduces the complexity but limits the ability to model global dependencies. SW-MSA addresses this limitation by alternating between regular and shifted windows to enhance long-range dependency modeling. However, due to the sparse spatial information in spectral images, its efficiency is lower than that of methods focusing on spectral correlation.

SMSA treats each spectral feature map as a token and computes attention along the spectral dimension, effectively capturing spectral dependencies. The computational complexity of W-MSA, SW-MSA, and SMSA scales linearly with spatial size, i.e., \( O(HW) \), which is significantly lower than the quadratic complexity \( O(HW^2) \) of G-MSA. As summarized in \autoref{tab:stage_number_attention_comparison}, although SMSA exhibits slightly higher complexity than W-MSA and SW-MSA, it achieves better reconstruction performance, demonstrating a more favorable trade-off between efficiency and accuracy. This indicates that SMSA can more effectively exploit spectral features and is better suited for spectral image colorization than mechanisms focusing solely on spatial dimensions.

\begin{figure*}[htbp]
    \centering
    \includegraphics[scale=0.56, trim=0 0 0 0, clip]{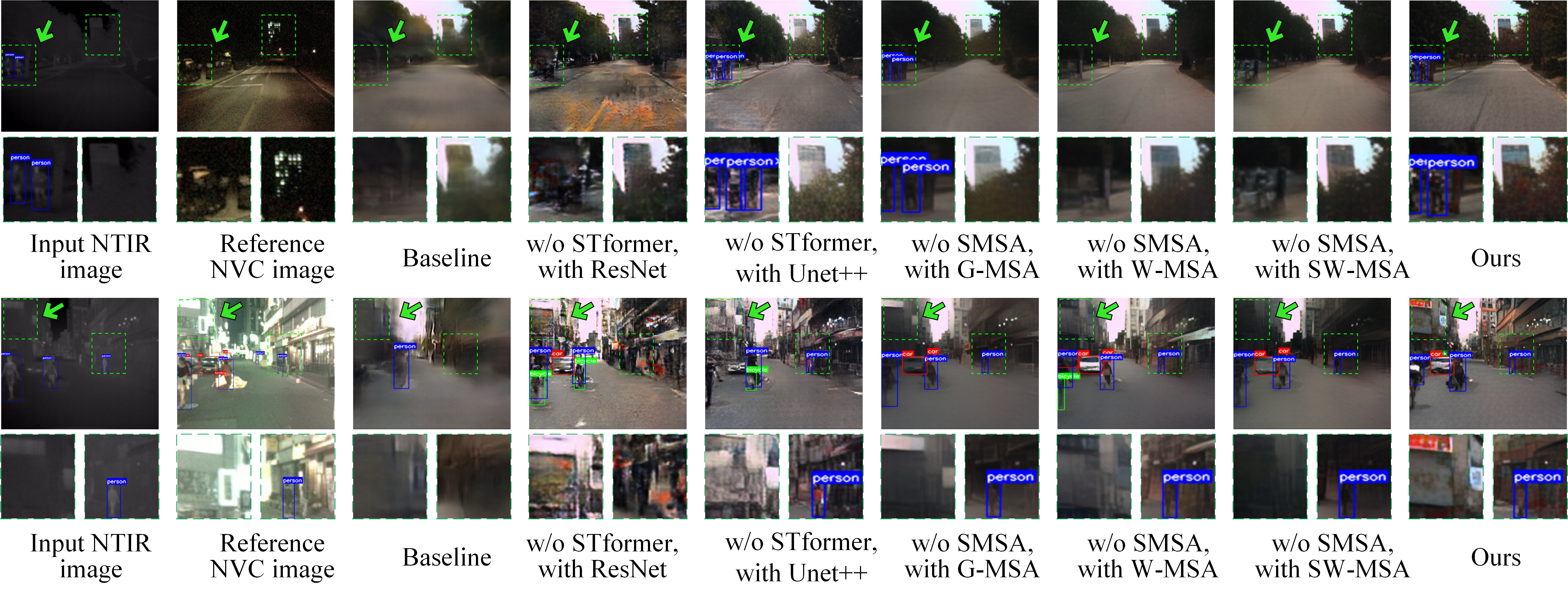}
    \caption{The ablation analysis of pedestrian detection using various YOLOv7-based network variants on the KAIST dataset is presented, where pedestrians are indicated by blue bounding boxes and vehicles are denoted by red ones. The 'w/o' indicates the removal of the module, while 'with' denotes the replacement of the original module.}
    \label{fig:STformer_ablation}
\end{figure*}

\textbf{Analysis of STformer:} The overall architecture of the MTSIC generator is shown in \autoref{fig:Overall architecture}(a), where ConvNeXt\cite{[66]} is used for pretraining to extract spatial features. To establish a baseline, STformer is removed, leaving only the ConvNeXt backbone. To evaluate STformer's effectiveness, ResNet~\cite{ResNet} and UNet++~\cite{Unet++} replace it, while the rest of the architecture remains unchanged. Performance comparisons, presented in \autoref{tab:Ablation_STformer_MSWB_and_loss}, show that removing or replacing STformer results in performance degradation, emphasizing its ability to integrate spatial and spectral features. As illustrated in \autoref{fig:STformer_ablation}, without STformer, small objects are often missed, and boundary ambiguity arises, further confirming its superiority in feature representation and detail preservation.

\begin{figure*}[htbp]
    \centering
    \includegraphics[scale=0.45, trim=0 0 0 0, clip]{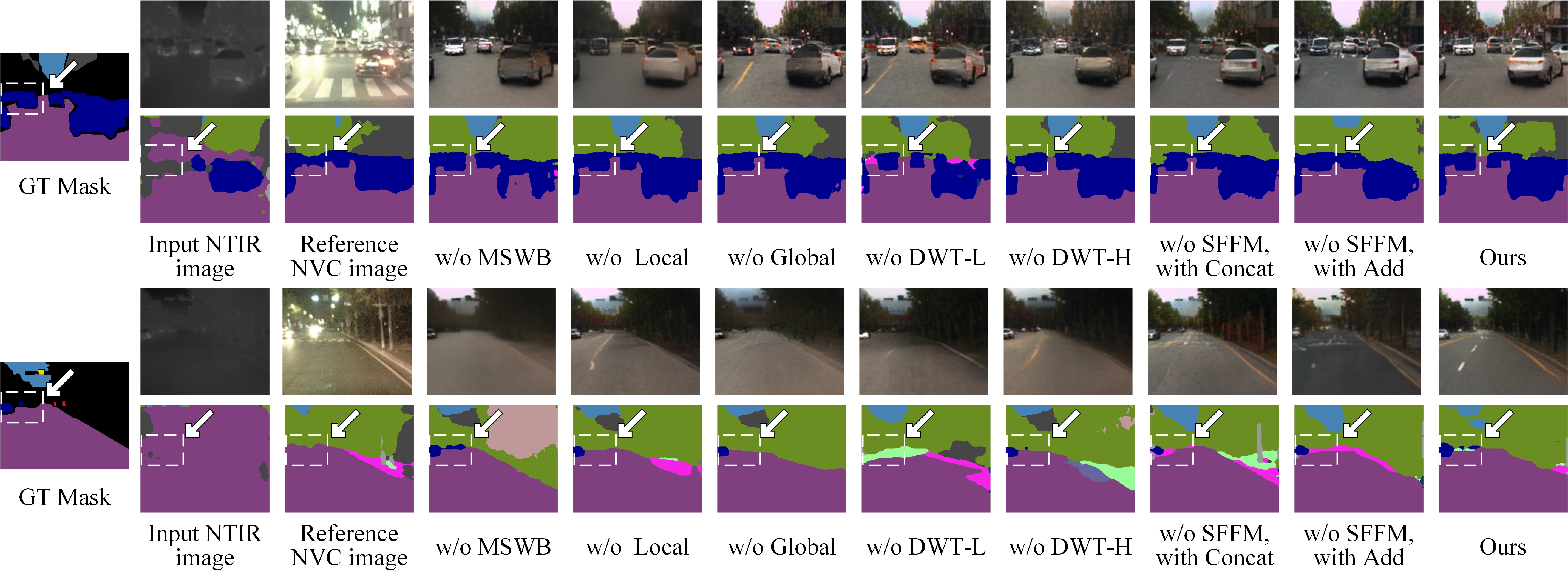}
    \caption{An ablation study was conducted on the MSWB module, where “w/o” indicates the module is removed, and “with” denotes that the module is used as a replacement.}
    \label{fig:MSWB_Abaltion}
\end{figure*}

\begin{figure}[htbp]
    \centering
    \includegraphics[scale=0.50, trim=0 0 0 0, clip]{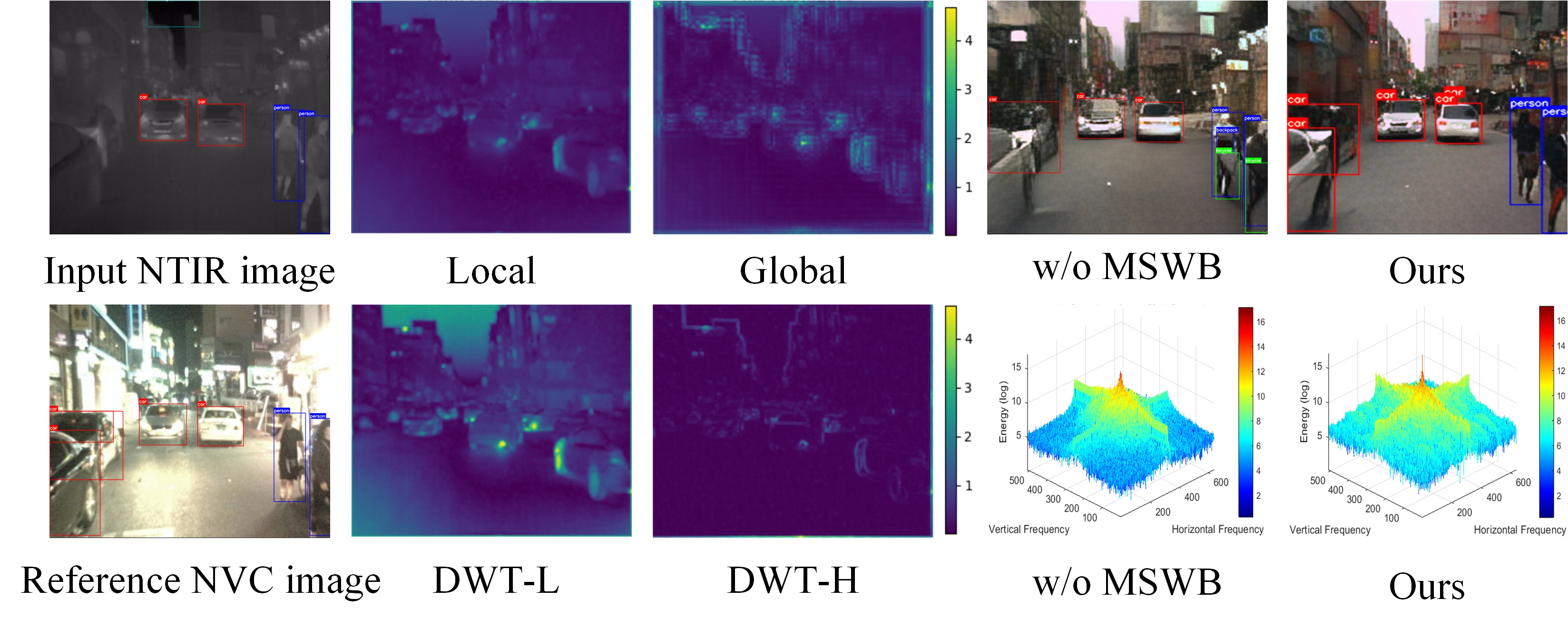}
    \caption{Feature map visualizations of each branch in the MSWB module.}
    \label{fig:MSWB_feature_map_Ablation}
\end{figure}

\begin{figure}[htbp]
    \centering
    \includegraphics[scale=0.20, trim=0 0 0 0, clip]{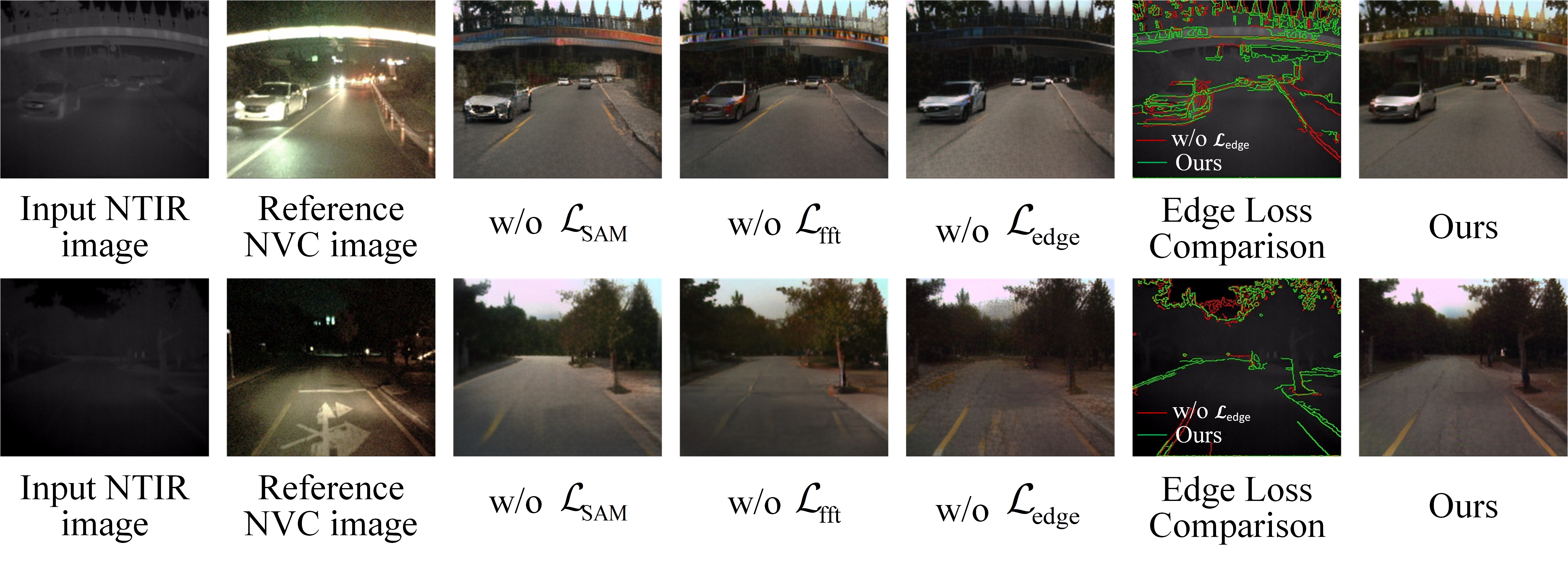}
    \caption{Loss ablation study on the nighttime KAIST dataset.}
    \label{fig:Loss_ablation}
\end{figure}

\begin{table}[htbp]
  \centering
  \setlength{\tabcolsep}{2.1pt} 
  \caption{Ablation study on the effectiveness of STformer, MSWB, and loss components. 'w/o' indicates the removal of the module, while 'with' denotes the replacement of the original module.}
  \begin{tabular}{
    >{\raggedright\arraybackslash}p{2.1cm}@{} 
    >{\raggedright\arraybackslash}p{3.2cm} 
    ccc
  }
    \toprule
    \textbf{Comparison} & \textbf{Variant} & \textbf{Precision} & \textbf{Recall} & \textbf{mAP50} \\
    \midrule
    \multirow{4}{*}{\makecell[l]{STformer\\Analysis}} 
      & Baseline & 21.6  & 17.3  & 11.2 \\
      & \makebox[3.2cm][l]{w/o STformer, with ResNet} & 34.2  & 24.7  & 20.9 \\
      & \makebox[3.2cm][l]{w/o STformer, with UNet++} & 52.7  & 35.9  & 36.3 \\
      & Ours & \textbf{64.2} & \textbf{41.3} & \textbf{40.8} \\
    \midrule
    \multirow{8}{*}{\makecell[l]{MSWB\\Analysis}} 
      & w/o MSWB & 53.7  & 31.0  & 32.9 \\
      & w/o Global & 61.2  & 36.7  & 38.4 \\
      & w/o Local & 60.7  & 33.5  & 36.3 \\
      & w/o DWT-L & 59.1  & 40.5  & 38.1 \\
      & w/o DWT-H & 55.6  & 36.7  & 37.6 \\
      & \makebox[3.2cm][l]{w/o SFFM, with Concat} & 62.3  & 40.1  & 39.3 \\
      & \makebox[3.2cm][l]{w/o SFFM, with Add} & 59.6  & 39.7  & 38.9 \\
      & Ours (with MSWB) & \textbf{64.2} & \textbf{41.3} & \textbf{40.8} \\
    \midrule
    \multirow{4}{*}{\makecell[l]{Loss\\Analysis}} 
      & w/o $\mathcal{L}_{\text{SAM}}$ & 63.5  & 40.2  & 39.7 \\
      & w/o $\mathcal{L}_{\text{fft}}$ & 61.9  & 38.7  & 37.3 \\
      & w/o $\mathcal{L}_{\text{edge}}$ & 63.4  & 39.6  & 38.1 \\
      & Ours & \textbf{64.2} & \textbf{41.3} & \textbf{40.8} \\
    \bottomrule
  \end{tabular}
  \label{tab:Ablation_STformer_MSWB_and_loss}
\end{table}

\textbf{Analysis of the MSWB:} \autoref{tab:Ablation_STformer_MSWB_and_loss} presents the impact of progressively removing sub-modules from MSWB. "Global" and "Local" represent the global and local branches, while "DWT-L" and "DWT-H" denote low- and high-frequency features from DWT decomposition. "SFFM" refers to the spatial-frequency feature fusion module. To validate SFFM's effectiveness, we compared performance with SFFM removed, and frequency-domain and spatial-domain features fused by concatenation ("w/o SFFM, with Concat") or element-wise addition ("w/o SFFM, with Add"). Note that spatial-scale features are downsampled according to the wavelet transform scaling ratio to achieve fusion between spatial and frequency domains. Moreover, if any branch of SFFM is removed, its corresponding SFFM alignment module is also eliminated and replaced with an Add operation.

\autoref{fig:MSWB_Abaltion} illustrates the impact of removing sub-modules, showing a performance drop in all cases, emphasizing the importance of each component. Replacing SFFM with Concat or Add caused significant degradation on the KAIST dataset due to the semantic gap between frequency-domain and spatial-domain features, which simple fusion methods fail to align effectively. The proposed fusion strategy, however, better captures semantic correlations, improving semantic consistency and segmentation accuracy in cross-domain image colorization.

\autoref{fig:MSWB_feature_map_Ablation} presents feature maps for each submodule, demonstrating their roles in enhancing feature representation. The Local branch captures fine details (e.g., vehicles), while the Global branch targets larger regions. DWT-H highlights local edge information, and DWT-L improves global perception. However, the Local branch is less effective for complex textures (e.g., buildings). Removing any branch causes structural distortion in reconstructed images and reduces segmentation performance. Compared to removing the MSWB module (“w/o MSWB”), incorporating it (“Ours”) enhances frequency-domain energy and significantly improves spatial visualization, as seen in \autoref{fig:MSWB_feature_map_Ablation}.

\textbf{Loss Analysis:} As described in Section 3, we propose a composite loss function combining $\mathcal{L}_{\text{SAM}}$, $\mathcal{L}_{\text{fft}}$, and $\mathcal{L}_{\text{edge}}$
, improving upon the strategy of Kuang et al.~\cite{[16]}. To evaluate the contribution of each component, we conducted three ablation studies on the KAIST dataset, each excluding one loss term while keeping the network architecture and training data unchanged. The results in \autoref{tab:Ablation_STformer_MSWB_and_loss} and \autoref{fig:Loss_ablation} demonstrate that removing $\mathcal{L}_{\text{fft}}$ leads to the most significant performance degradation, highlighting the importance of spatial-frequency optimization. Omitting $\mathcal{L}_{\text{SAM}}$ results in a slight decline in Precision and Recall, suggesting its role in enhancing spectral consistency through angular similarity; moreover, including $\mathcal{L}_{\text{SAM}}$ results in more natural-looking images. Although $\mathcal{L}_{\text{edge}}$ has limited impact on detection performance, it aids in preserving edge structures and semantic consistency. In \autoref{fig:Loss_ablation}, contours with and without edge loss are overlaid on the input image. The version with edge loss (“Ours”) better preserves contours and and mitigates boundary artifacts. 

\section{Conclusion}

This study proposes a novel method for multi-band infrared spectral image colorization, named MTSIC. It exploits the spatial sparsity and spectral self-similarity of multi-band infrared images. Each spectral feature map is treated as a token for self-attention within the SARB unit. Multiple SARB units are stacked to form the STformer, and several STformer modules are cascaded to construct the MTSIC framework. A multi-stage learning strategy is employed to progressively refine the reconstruction, from coarse to fine. Within the STformer, an MSWB module applies the Haar wavelet transform for frequency-domain feature mapping, while the SFFM module integrates low-frequency and global features with high-frequency and local features, bridging the semantic gap between the frequency and spatial domains. 

Experimental results demonstrate that the proposed method significantly improves reconstruction quality. This work addresses challenges in infrared image colorization by leveraging spectral properties and encourages future research into combining Transformer architectures with wavelet transforms for infrared hyperspectral image colorization, opening new directions for further exploration. 



%





\ifCLASSOPTIONcaptionsoff
  \newpage
\fi





\bibliographystyle{IEEEtran}
\bibliography{IEEEabrv,Bibliography}


\vfill

\end{document}